\documentclass{aa}  

\usepackage{graphicx}
\usepackage{txfonts}
\usepackage{amsmath}	
\usepackage{amssymb}	
\usepackage{xcolor}
\usepackage[colorlinks=true,citecolor=blue,linkcolor=blue]{hyperref}
\usepackage{caption}
\usepackage{subcaption}
\usepackage{multirow}

\newcommand{\ur}[1]{\mathrm{#1}}
\renewcommand{\v}[1]{\boldsymbol{#1}}
\renewcommand{\d}{\ur{d}}

\newcommand{\kms}{\,\mathrm{km}\,\mathrm{s}^{-1}}
\newcommand{\Msyr}{\,\mathrm{M}_\odot\,\mathrm{yr}^{-1}}
\newcommand{\ergs}{\,\mathrm{erg}\,\mathrm{s}^{-1}}

\newcommand{\pccm}{\,\mathrm{cm}^{-3}}
\newcommand{\Msun}{\,\mathrm{M}_\odot}

\newcommand{\Lw}{L_\mathrm{w}}

\newcommand{\Rc}{R_\mathrm{c}}

\newcommand{\Bm}{$\left<B\right>$}

\begin{document}

   \title{Stellar-wind feedback and magnetic fields around young compact star clusters: 3D MHD simulations}
   \titlerunning{Wind feedback and magnetic field around young star clusters}
   \authorrunning{L. H\"arer et al.}

   \subtitle{}

   \author{L. Härer\thanks{\email{lucia.haerer@mpi-hd.mpg.de}}, 
            T. Vieu, and B. Reville}

   \institute{Max-Planck-Institut f\"ur Kernphysik, Saupfercheckweg 1, 69117 Heidelberg, Germany}
   \date{Received 06/02/2025; accepted 24/03/2025}


  \abstract
   {The environments of young star clusters are shaped by the interactions of the powerful winds of massive stars, and their feedback on the cluster birth cloud. Several such clusters show diffuse $\gamma$-ray emission on the degree scale, which hints at ongoing particle acceleration.}
   {To date, particle acceleration and transport in star-cluster environments are not well understood. A characterisation of magnetic fields and flow structures is necessary to progress toward physical models. Previous work has largely focused on 100\,pc scale feedback or detailed modelling of wind interaction of just a few stars. We aim to bridge this gap. We focus in particular on compact clusters to study collective effects arising from stellar-wind interaction. Objects in this class include Westerlund~1 and R136.}
   {We perform 3D ideal-MHD simulations of compact, young, massive star clusters. Stellar winds are injected kinetically for 46 individual very massive stars ($M>40\,\Msun$), distributed in a spherical region of radius ${\leq}$ 1\,pc. We include a sub-population of five magnetic stars with increased dipole field strengths of 0.1--1\,kG. We study the evolving superbubble over several 100\,kyrs.}
   {The bulk flow and magnetic fields show an intricate, non-uniform morphology, which is critically impacted by the relative position of individual stars. The cluster wind terminates in a strong shock, which is non-spherical and, like the flow, has non-uniform properties. The magnetic field is both composed of highly tangled sections and coherent quasi-radial field-line bundles. Steep particle spectra in the TeV domain arise naturally from the variation of magnetic field magnitude over the cluster-wind termination shock. This finding is consistent with $\gamma$-ray observations. The scenario of PeV particle acceleration at the cluster-wind termination shock is deemed unlikely.}
  {}
   

   \maketitle
%

\section{Introduction}
\label{sec:intro}

Massive stars feed back onto their environment through powerful winds and ionising radiation. The majority of massive stars live in young star clusters \citep{higdon05}. The collective feedback from such clusters shapes their local environment and impacts galaxy evolution \citep[see][]{schinnerer24}. Stellar winds in particular can excavate ambient medium, creating so-called superbubbles, which are filled with hot, tenuous, turbulent gas and extend over a scale of 100\,pc \citep[see][]{chu08}. Superbubbles have been observed, for example, in the 30 Doradus region in the Large Magellanic Cloud \citep[e.g.][]{Townsley2006}. The impact of wind feedback on a galactic scale is strikingly revealed by recent observations with the James Webb Space Telescope \citep[][and other works published in the same issue]{lee23}. Young star clusters present a diverse source class. They can differ by their extension, stellar composition, and environment. Stellar-wind interaction and feedback depend on these parameters. Compact, massive clusters in particular produce powerful winds over a scale of several 10\,pc. Such clusters include for example Westerlund~1 and R136 \citep[see][]{portegieszwart10}. 

In recent years, regions harbouring massive stars have been found to emit $\gamma$-rays. In the TeV domain, this includes Westerlund~1 \citep{hess22-wd1}, Westerlund~2 \citep{hess11-wd2}, the Cygnus region \citep{lhaaso23-cygnus}, W43 \citep{lhaaso24-w43}, R136, and 30~Doradius~C \citep{hess24-lmc}. Emission from these objects can extend to photon energies ${\gtrsim}100\,$TeV, typically with steep, curved spectra. In particular the exceptional detection of PeV photons from the Cygnus region raises the question as to which role star-cluster environments play in the Galactic cosmic-ray ecosystem. The isolated-supernovae standard paradigm for the origin of Galactic cosmic rays is challenged by several observational facts, such as the anomalous $^{22}\ur{Ne}/^{20}\ur{Ne}$ ratio \citep[see][]{gabici19, gupta20}. This measurement and the recent $\gamma$-ray detections, among other considerations, motivate the investigation of non-thermal processes in young star-cluster environments. Furthermore, the galactic-scale feedback from young star clusters impacts particle transport in the galactic disc and escape into the halo. In turn, the physics of the local medium, the dynamics of giant molecular clouds, and galaxy evolution as a whole are affected by escape of non-thermal particles from their productions sites and transport within the Galaxy. 

Non-thermal processes in star-cluster environments are sensitive to stellar-wind feedback across several scales, from the sub-parsec range to the scale of the superbubble (100\,pc). A number of scenarios for the production of non-thermal particles coexist \citep[e.g.][]{bykov92, klepach00,ferrand10,gupta20,morlino21,vieu22a,vieu23}. The parameter space of these models is not well constrained, due to the large number of free parameters resulting from the complexity of star-cluster environments and the difficulty of constraining many of these parameters from observations. This concerns in particular the cluster-wind termination shock, which has been purported to be a favourable site of particle acceleration \citep[e.g.][]{morlino21}. Analytic models for wind interaction in star clusters have been developed in idealised spherically symmetric cases \citep{chevalier85,canto00}. Such models do not capture the 3D morphology and, in particular asymmetries introduced by powerful individual stars, such as Wolf-Rayet stars. In addition, 1D analytical models do not account for how the magnetic field is shaped by stellar-wind interaction. However, the magnetic fields' details are of prime importance for the acceleration and confinement of particles in a star-cluster environment. These issues motivate detailed studies of stellar-wind interaction using numerical simulations. Star cluster feedback has been investigated in a variety of works over the past decades \citep[e.g.][]{krause2013,Rogers2013,gupta18, elbadry2019,Lancaster2021}. These studies emphasise the superbubble-scale and feedback onto molecular clouds and the interstellar medium. They therefore employ simple wind injection schemes, such as the injection of thermal energy. To understand the flow, shock conditions, and magnetic field geometry in the depths of the superbubble, however, requires kinetic injection of individual winds. In particular, the cluster core must be resolved at sub-parsec scales. This is a challenging task which has only been attempted in a handful of recent works \citep{badmaev22, badmaev23, vieu24-core, vieu24-cyg}. A comprehensive MHD simulation study including sub-particle to superbubble scales is lacking thus far. 

The present work aims at filling this gap. We place particular emphasis on potential particle acceleration sites, such as the cluster-wind termination shock, in order to discuss implications for $\gamma$-ray emission. Since our aim is to study collective effects of stellar-wind interaction, we focus on compact clusters. In this context, compact clusters are defined as those in which collective effects are strong enough to produce a cluster-wind termination shock. In loose clusters, such as Cygnus OB2, this is not the case \citep[see][]{vieu24-cyg}. The present work is structured as follows: Sect.~\ref{sec:setup} describes how stars and the cluster are modelled. In Sect.~\ref{sec:evo} and \ref{sec:b}, we describe results concerning the hydrodynamical behaviour and the magnetic field, respectively. Based on our findings, we then discuss particle acceleration and propagation in star-cluster environments in Sect.~\ref{sec:disc} and conclude with Sect.~\ref{sec:concl}. 

\section{Setup}
\label{sec:setup}

Our aim is to investigate stellar-wind interactions around prototypical compact, young, massive clusters to gain insight into the nature of collective effects therein and how such regions may accelerate particles to high energies. The complementary case of loose clusters was discussed in \citet{vieu24-cyg} inspired by the example of the Cygnus star-forming region. We begin with a description of the numerical setup in Sect.~\ref{sec:num} and then introduce how the star cluster and ambient medium are modelled in Sect.~\ref{sec:amb}--\ref{sec:runs}. All parameters of our model star cluster are summarised in Tab.~\ref{tab:stars}.

\subsection{Numerical setup, grid, and outer boundary}
\label{sec:num}

We utilise the publicly available finite-volume code \texttt{PLUTO} \citep{mignone07}, version 4.4-patch2, to solve the ideal-MHD equations:
\begin{align}
    &\frac{\partial\rho}{\partial{t}}=-\nabla\cdot\left(\rho\v{u}\right) \, ,\\
    &\rho \left(\frac{\partial\v{u}}{\partial t} + (\v{u}\cdot \nabla) \v{u} \right) = \nabla\cdot\left(p+\frac{B^2}{8\pi}\right) + \frac{1}{4\pi}\left(\v{B}\cdot\nabla\right)\v{B} \,  ,\\  
    &\frac{\partial\v{B}}{\partial{t}}=\nabla\times\left(\v{u}\times{B}\right) \, ,
\end{align}
where $\rho$ is the mass density, $\v{u}$ the velocity, $p$ the pressure, and $\v{B}$ the magnetic field vector. We adopt an ideal equation of state with an adiabatic index of 5/3. We choose the \texttt{TVDLF} Riemann solver, second-order Runge-Kutta time-stepping, and linear spatial reconstruction for second-order accuracy in space. The divergence-cleaning algorithm (\texttt{div\_cleaning}) is used to enforce the $\nabla\cdot \mathbf{B} = 0$ condition over the course of the simulation. We switch to the entropy equation \citep{balsaraspicer} if the flow speed exceeds 2000\,$\kms$ and the sonic Mach number is ${>}3$. This setting is more restrictive than the one employed for the \texttt{selective} entropy switch implemented in \texttt{PLUTO}. Our setting ensures that the deviation from global energy conservation due to the use of the entropy equation stays at the percent level (see appendix~\ref{app:1d}). The boundary condition on the external domain is set to \texttt{outflow}. Further description of the \texttt{PLUTO} code can be found in the User's Guide\footnote{\url{http://plutocode.ph.unito.it/userguide.pdf}}. 

We use a non-uniform Cartesian grid to facilitate both the implementation of individual stellar winds in the core, and the extension of the grid to the scale of the superbubble. In a cube ${\pm}1\,$pc from the centre of the domain, the grid has a uniform resolution of 0.008\,pc per cell. At $|x|$, $|y|$, $|z|>1\,$pc, the \texttt{PLUTO} stretched grid function is employed to progressively decrease the resolution down to 0.96\,pc, individually along each axis. A side effect of this setup is the presence of rectangular-cuboid cells along the coordinate axes at large distances from the core. We verified that these rectangular cells do not significantly affect the flow pattern, by performing a test run of a cluster rotated by 45$^\circ$ along the $x$, $y$, and $z$ axes. All simulations have a total of $512^3$ cells corresponding to a physical width of the domain of 56\,pc. 

Our setup is tailored to study the cluster wind termination shock region over several hundreds of kyrs. We chose to invest resources to achieve high cluster core resolution over a large simulation domain and long timescale rather than include additional physical processes, such as radiative losses and thermal conduction. On the investigated timescales, radiative losses are not expected to have a direct impact on the low-density superbubble interior \citep[see][]{weaver77,elbadry2019}. Previous work on star-cluster wind-interaction has explored the effects of thermal conduction \citep[][]{badmaev22,badmaev23} and found it to smooth structures in the core but not introduce major morphological changes. Some of the impacts of physical processes left out here can be anticipated from simulations of wind interaction at smaller scales \citep[e.g.][and references therein]{mackey25}.

\subsection{Ambient medium}
\label{sec:amb}

The ambient medium is initialised as a homogeneous medium with an ideal equation of state,
\begin{equation}
p = c_\ur{s}^2 \frac{\rho}{\gamma} \, ,
\end{equation}
where $\gamma = 5/3$ is the adiabatic index . The sound speed is 
\begin{equation}
c_\ur{s} = 15 \left( \frac{T}{10^4\,\ur{K}} \right)^{0.5} \left( \frac{\gamma}{5/3} \right)^{0.5} \left( \frac{\mu}{0.61} \right)^{0.5} \,\kms  \,
\label{eq:cs}
\end{equation}
where $10^4\,$K is the temperature assumed for all runs. We assume a fully ionised gas at solar metallicity, resulting in a mean molecular weight of $\mu = 0.61$. The simulation box is permeated by a uniform 3.5\,$\mu$G magnetic field along the $x$-direction, which is in the range that is typically expected in the interstellar medium \citep[e.g.][]{unger24}. The ambient density is $100\pccm$, which is a typical value for giant molecular clouds, but higher than the average density of the interstellar medium. The density mainly acts as a scaling factor for the size of the superbubble. The choice of a high value is necessary to confine the superbubble in the simulation box over the run time of the simulation.

\subsection{Star cluster and winds}
\label{sec:stars-pars}

We generate a synthetic population of 46 stars with masses above $40\Msun$ and simulate their wind feedback for a total time of 390\,kyr. Stellar masses are distributed according to the initial mass function $\d N \sim M^{-2.3} \d M $ \citep[][]{kroupa02}. This represents the high-mass end of a star cluster with a total number of 500 massive stars ($M>8\Msun$) and total mass of $3.5\times 10^{4}\Msun$ above $0.4\Msun$. 

All stars are initialised on the main sequence. Stellar positions are drawn randomly from a uniform distribution within a sphere of radius $\Rc$. Stellar and wind parameters are assigned based on the mass of the model stars. We initialise the individual stellar winds at terminal velocity, since the smallest simulated scale is 0.008\,pc, which greatly exceeds the largest possible stellar radius. For the terminal wind velocity, $u_\infty$, and mass-loss rate, we adopt relations by \citet{seo18}. These relations are based on grid models for stellar evolution by \citet{ekstrom12} and mass-loss predictions by \citet{vink01}. The terminal wind velocity is set to $u_\infty = 2.6u_\ur{esc}$ for O stars above the bi-stability jump \citep{vink01}. We only consider stars with $M>40\Msun$, which are well above the bi-stability jump. The mass-loss rate of an O star on the main sequence is (in $\Msyr$),
\begin{equation}
    \log_{10} \dot{M} = -3.38\times(\log_{10}M)^2+14.59\times\log_{10}M-20.84 \, ,
\end{equation}
and the wind power is (in $\kms$)
\begin{equation}
    \log_{10} L_\ur{w} = -3.38\times\left(\log_{10}M\right)^2+14.77\times\log_{10}M+21.21 \, .
\end{equation}
The wind velocity is calculated from these two parameters. The mass is in units of solar masses ($\Msun$). In the main sequence phase, the wind power of our model cluster is $\Lw = 0.5\dot{M}u_\infty^2 = 3\times 10^{38}\ergs$. The total power of the excluded stars with $M<40\,\Msun$ is subdominant, $4\times10^{37}\ergs$. 

All parameters are kept fixed for the first 200\,kyr of the simulation, after which the most massive stars enter a Wolf-Rayet (WR) phase. WR stars have exceptionally powerful winds, and in many clusters are expected to dominate the wind power \citep[e.g.\ in Cygnus OB2, see][]{vieu24-cyg}. The mass-loss rate and the terminal wind-velocity in the WR phase depend on stellar mass and WR-type \citep[][]{hamann19,sander19}. Here, we apply a simple uniform prescription: we increase the main-sequence mass-loss rate by a factor of ten. The first massive star enters the WR phase at 200\,kyr. Less massive stars enter the WR phase with a time-delay according to their lifetime, following the prescription of \citet{zakhozhay13}, whose work is based on \citealt{schaller92}. While the onset-time of the WR phase (200\,kyr) is much shorter than the expected main sequence life-time of a massive star, we will show that it is sufficient for the simulation to reach a quasi-stationary state, in other words, no further major morphological changes occur in the flow and magnetic field pattern. Between simulation time 200\,kyr and 390\,kyr, a total of five stars enter the WR stage.

Our wind model is qualitative, yet is sufficient to explore the phenomenology of wind flows and magnetic fields in and around compact star clusters, in particular with the aim to better understand the physics of wind interactions. More detailed stellar evolution, stellar motion, binary population, supernovae or red supergiants models present opportunities for future work.

\subsection{Stellar radius, rotation, and temperature}

To implement the magnetic field and the initial conditions in the vicinity of the stars, we require expressions for stellar radii, temperatures, and rotational velocities. We set the former two values based on \citet{eker18}, employing their mass-temperature and mass-luminosity relations,
\begin{align}
\log_{10} T (M) &= -0.170\times\left(\log_{10}M\right)^2 + 0.888\times \log_{10}M + 3.671 \, , \\
\log_{10} L (M) &= \left(2.865\times\log_{10}M + 1.105 \right) \, .
\end{align}
where $M$ is given in $\Msun$, $L$ in $\ur{L}_\odot$, and $T$ in K. The relation for $T$ is also used to set the wind sound speed, $c_\ur{s}=31\mbox{--}36\,\kms$ for the chosen mass range, set according to Eq.~\ref{eq:cs} with $\gamma=1$ for isothermal winds. Since there is no clear scaling with $M$, the value of $u_\ur{rot}$ is fixed at $300\kms$ for all stars \citep[cf.][]{grunhut17}, which amounts to 30-50\% of the critical velocity for the stellar masses in our cluster. 

The threshold mass of the empirical relations given by \citet{eker18} is $31\Msun$, due to limited statistics at higher masses. A comparison between the O star calibration by \citet{martins05}, which is based on an empirical approach and models allowing for states outside of local thermodynamic equilibrium (non-LTE models), yields an overestimation of $T$ and $R_\star$ by 11\% and 14\% when using the \citet{eker18} relations at $58\Msun$. Note that our model does not require precise values for $T$ and $R_\star$. The former value influences exclusively the sound speed inside the stellar-wind initialisation region and is only weakly dependent on $T$. The stellar radius, $R_\star$, merely acts as a scaling factor on $B$, as described in Sect.~\ref{sec:b-setup}.

\subsection{Stellar magnetic fields}
\label{sec:b-setup}

The strength and morphology of the magnetic fields of massive stars are not well constrained, but existing data are consistent with a dipolar field structure and a bimodal field-strength distribution \citep[e.g.][]{grunhut17}, where ${\lesssim}10\%$ of massive stars have a dipolar surface field strength averaging at ${\sim}1\,$kG at the stellar surface, while the remainder of stars have no detectable fields. We set the field strength of these non-magnetic stars to a fiducial value of 10\,G, which would be below the detection threshold. In interplay with stellar rotation, the dipolar field results in a Parker spiral \citep{parker58} beyond the Alfv\'en radius, where the radial and azimuthal components of the field are given as,
\begin{align}
	B_\mathrm{r} \,&= B_0(\theta) \left(\frac{R_\star}{r} \right)^2 \, , \label{eq:parker1}\\
	B_\varphi &= B_0(\theta) \frac{u_\mathrm{rot}}{u_\mathrm{w}} \sin \theta \frac{R_\star}{r} \left( 1 - \frac{R_\star}{r}\right) \, ,
	\label{eq:parker2}
\end{align}
assuming the magnetic axis aligns with the axis of rotation. The expressions above depend on $B_0$, the surface magnetic field at the stellar radius, $R_\star$, the ratio of the rotational velocity at the equator $u_\ur{rot}$ to the wind velocity $u_\ur{w}$, the polar angle, $\theta$, and the radius, $r$. At the resolution of the simulations presented in this work, the dependence of the surface field on the polar angle, $\theta$, can be approximated as a split monopole. We include a scaling function to smooth the jump at the stellar equator, 
\begin{equation}
B_0(\theta) = B_\mathrm{s} \tanh\left(3\left(\frac{\pi}{2} - \theta\right)\right) \, .
\end{equation}

We caution the reader that our knowledge of stellar magnetism relies on spectroscopic measurements at the stellar surface. Close to the surface, the scaling of the magnetic field with $r$ does not necessarily follow Eq.~\ref{eq:parker1} and \ref{eq:parker2}. Naively, one would expect $B$ to follow a dipole scaling inside the Alfv\'en radius. Neither the Alfv\'en radius in the wind launching region nor the stellar radius corresponding to the measured $B_0$ are well constrained. This introduces significant uncertainty to the overall appropriate scaling. In addition, we have fixed $u_\ur{rot} = 300\,\kms$ for all stars. The known population of magnetic stars shows on average lower $u_\ur{rot}$ than non-magnetic stars. For magnetic stars, $u_\ur{rot}$ can be as low as a few to a few 10s of $\kms$.  However, the variation between individual stars is large and some stars have rotational velocities comparable to those typical for non-magnetic stars \cite[e.g.][]{grunhut17}. This impacts our setup only as far as it reduces $B_\varphi$, which is the relevant component at $r\gg R_\star$. Overall, our assumption on stellar magnetic fields is on the optimistic side.

\subsection{Initialisation of stellar winds}

At the position of each star, we implement an internal outflow boundary with a radius of five cells, $R_\ur{b}=0.04\,$pc, in which we fix the wind parameters and the magnetic field as described above. Inside the boundary, $\rho$ follows from mass-continuity, 
\begin{equation}
\dot{M} = 4\pi r^2 u_\mathrm{w} \rho (r) \, .
\label{eq:mass-cont}
\end{equation}  
Outside $R_\ur{b}$, parameters are scaled smoothly to their ambient values to avoid discontinuities at initialisation. This procedure was found to reduce artefacts introduced by the initialisation of spherical wind boundaries on a Cartesian grid and increase stability. The density is scaled up to the ambient density, $\rho_\ur{amb}$, with the function
\begin{equation}
    \lambda(r) = \left(1 + \frac{r-R_\mathrm{b}}{R_\mathrm{s}}\right)^\alpha \ , \ur{where} \
    R_\mathrm{s} = \sqrt{\frac{\dot{M}}{4\pi\rho_\ur{amb} u_\ur{w}}} \, .
\end{equation}
The parameter $\alpha\geq2$ is set by the requirement that the ambient density be reached at a radius $R_\ur{t}=2R_\ur{b}$ from each star. Scaling down the velocity and magnetic field as 
\begin{align}
u(r) &=  \frac{u_\ur{w}}{\lambda(r)} \, , \\
B_\varphi(r) &= B_\varphi(R_\mathrm{b}) \frac{R_\mathrm{b}}{r} \sqrt{\frac{1}{\lambda(r)}} \, 
\end{align}
ensures continuity and constant sonic and Alfv\'enic Mach numbers inside $R_\ur{t}$. In addition, $u_\mathrm{w}$ and $B_\varphi$ are scaled down to ${\lesssim}10\%$ of their value at the boundary.

\subsection{Simulation runs}
\label{sec:runs}

\begin{table} 
	\centering
	\begin{tabular}{c|c|l} 
		cluster & $R_\ur{c}$ & magn.~stars \\
		\hline
		\hline
		I & 0.6\,pc & 10\%, 1\,kG\\ 
		I (high $B$) & 0.6\,pc &20\%, 1\,kG\\  
		I (low $B$) & 0.6\,pc &10\%, 100\,G\\    
		II & 0.6\,pc & 10\%, 1\,kG \\ 
		III  & 1\,pc &  10\%, 1\,kG \\
		
	\end{tabular}
	\caption{Simulation parameters that differ between runs. A ``cluster'' is a specific spatial distribution of stars within a sphere of radius $R_\ur{c}$. We provide 3D renderings of clusters I--III in Fig.~\ref{fig:cluster3d} in the appendix. The stellar content of all clusters is the same and is summarised in Tab.~\ref{tab:stars}. The ``magn.\ stars'' column indicates the percentage of stars that has the magnetic field given in the column. For further details on the setup see Sect.~\ref{sec:setup}. }
	\label{tab:sims}
\end{table}

\begin{figure*}
    \centering
  	\includegraphics[width=\textwidth]{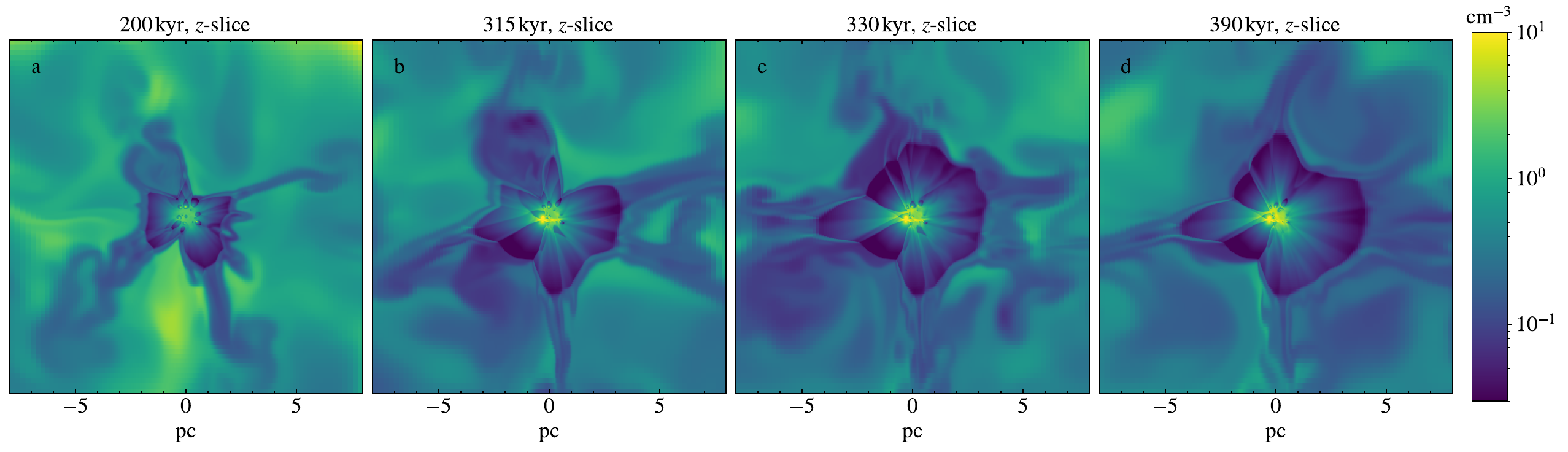}
  	\includegraphics[width=\textwidth]{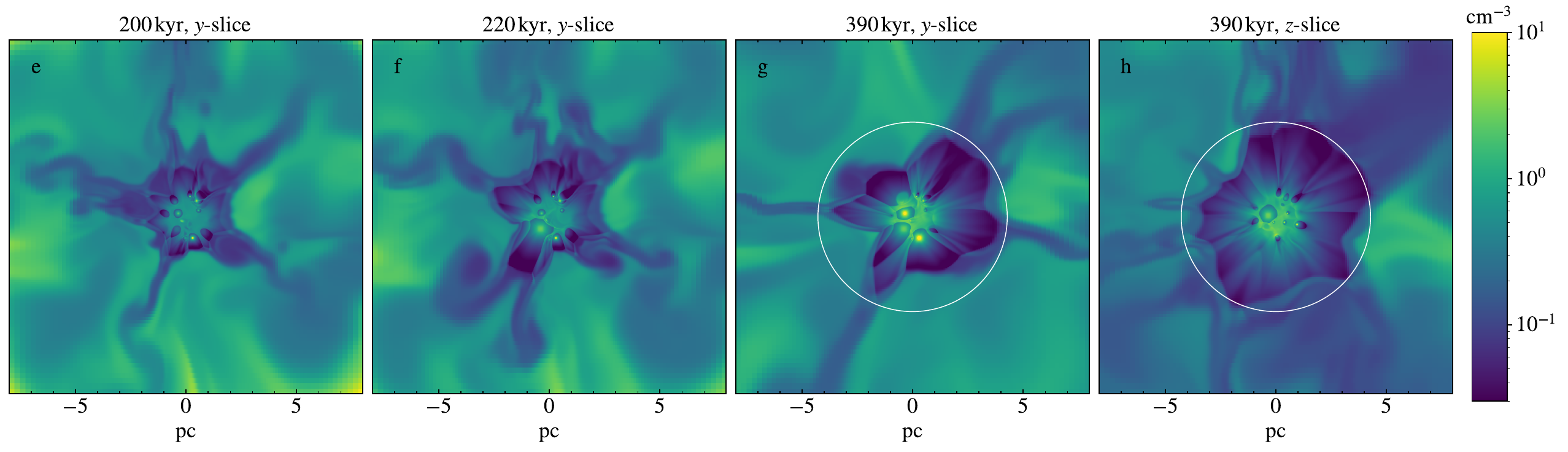}
  	\caption{Density slices for cluster I ($R_\ur{c}=0.6\,$pc, top) and cluster III ($R_\ur{c}=1\,$pc, bottom). The inner $\pm8$\,pc are shown, which excludes the superbubble shell. In the left-most panels (a and e), all stars are on the main sequence. The cluster wind is in a quasi-stationary state, slowly expanding over time but not changing geometry. At $t>200\,$kyr (panels b--d and f--h), the most massive stars consecutively evolve into Wolf-Rayet stars. The flow rearranges into a new quasi-stationary state within ${\sim}20$ kyr after each new Wolf-Rayet wind (see panels b--c and e--f). The wind termination shock has a complex, non-spherical geometry, which appears different in different slices (see panels g--h). The white circle in panels g--h indicates the shock radius predicted by 1D analytical theory \citep{weaver77}.}
  	\label{fig:rho-map}
\end{figure*}

We consider three clusters with identical stellar content but different spatial distributions, which we term clusters I--III. Table~\ref{tab:sims} shows an overview of all runs. The cluster radius is 0.6\,pc for clusters I and II and 1\,pc for cluster III. The high compactness of clusters I and II was chosen to maximise collective effects within the run time of the simulation. The radius of cluster III is representative of, for example, Westerlund~1 and 2 \citep[see][]{portegieszwart10}. As will be demonstrated with our results, cluster I/II and III have the same morphological qualities, as do test runs extending the radius to ${\sim}2$\,pc. The behaviour of the clusters studied here can therefore be considered prototypical for young compact clusters. Stellar and wind parameters were chosen as described in Sect.~\ref{sec:stars-pars} and are summarised in Tab.~\ref{tab:stars} in Appendix~\ref{app:tab-stars}. By default, five of the 46 stars in each cluster are magnetic with a surface field of 1\,kG. For cluster I, we investigate in addition cases of low and high $B$. In the low $B$ case, magnetic stars have a surface field strength of 100\,G instead of 1\,kG. In the high $B$ case, we double the number of magnetic stars to 10 stars, but keep the surface field strength of 1\,kG. Non-magnetic stars have a 10\,G field. Considering the literature, the high and low $B$ scenarios are atypical, but not implausible for individual clusters (see Sect.~\ref{sec:b-setup}).

\section{Bubble structure and evolution}
\label{sec:evo}

\begin{figure*}
	\begin{subfigure}{0.33\linewidth}
		\centering
		\includegraphics[width=\linewidth]{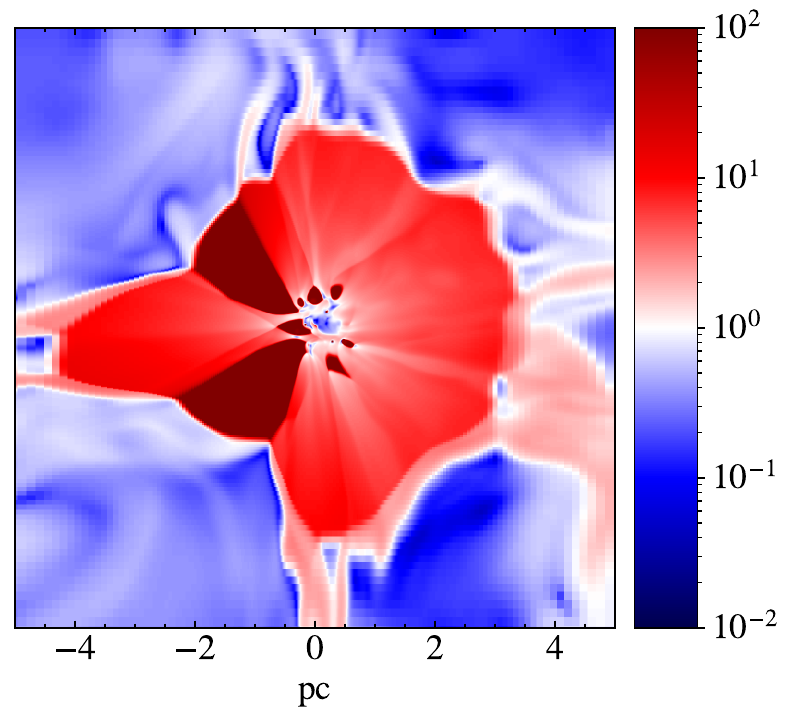}
		\caption{sonic Mach number}
		\label{fig:ms-map} 
	\end{subfigure}
	\begin{subfigure}{0.33\linewidth}
		\centering
		\includegraphics[width=0.97\linewidth]{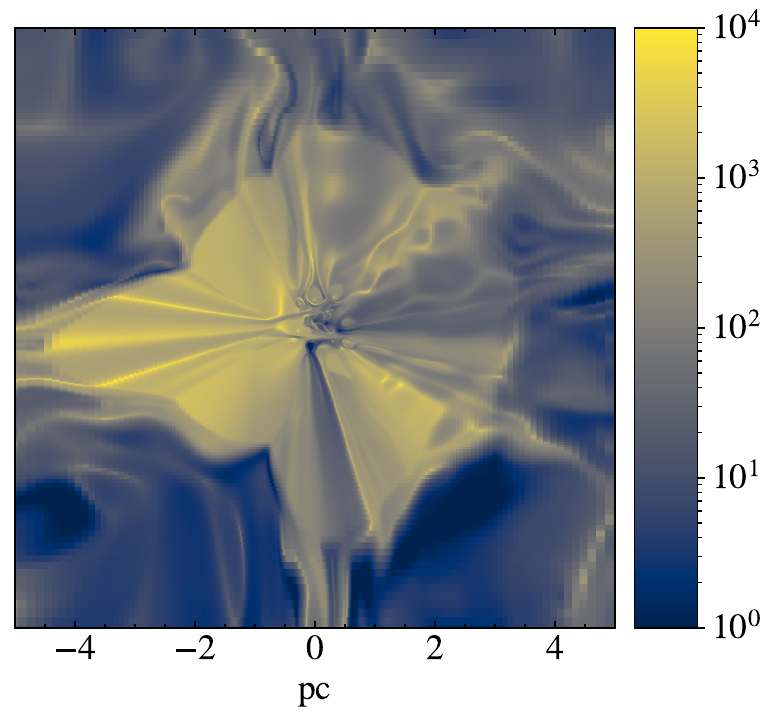}
		\caption{Alfv\'enic Mach number}
		\label{fig:ma-map} 
	\end{subfigure}
	\begin{subfigure}{0.33\linewidth}
		\centering
		\includegraphics[width=\linewidth]{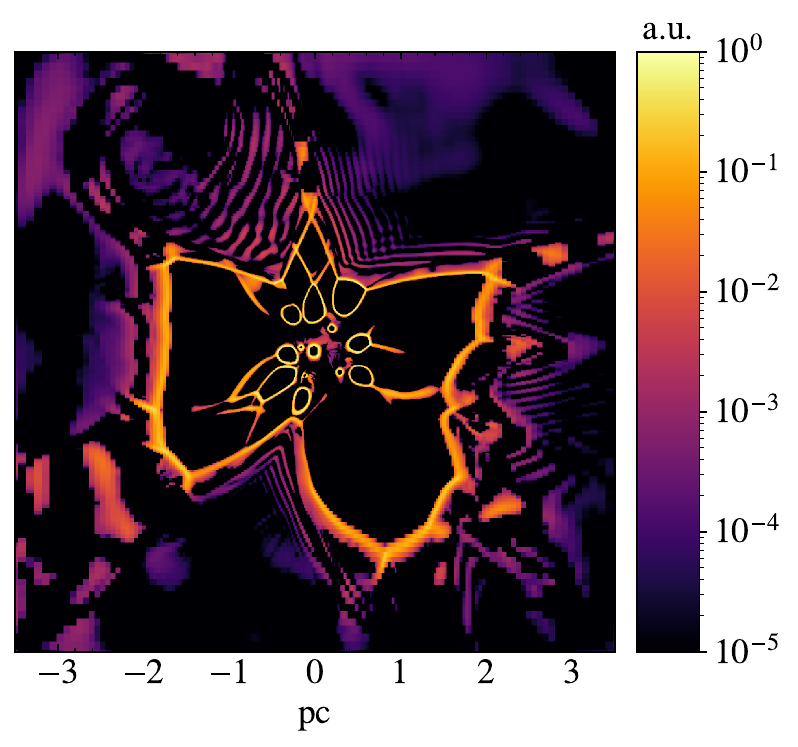}
		\caption{negative divergence of the flow} 
		\label{fig:divu-map} 
	\end{subfigure}
	\caption{Slices highlighting the properties of shocks in and around the star cluster. (a) Individual stellar winds reach Mach numbers of $M_\ur{S}\sim100$ and the cluster wind has $M_\ur{S}\sim10$. The transonic sheets, which are found downstream of the cluster-wind termination shock, have $M_\ur{S}\sim 1\mbox{--}2$. (b) The flow is highly super-Alfv\'enic, except in a small number of localised regions in the superbubble and cluster core. (c) The cluster-wind termination shock is compressive across its entire surface, as show by the negative divergence of the velocity field. Alternating compression and rarefaction zones inside the transonic sheets indicate a structure similar to that seen in the ``shock diamond'' phenomenon. The values are normalised to the maximum. Panels (a) and (b) show a $z$-axis slice of cluster I at 330\,kyr. Panel (c) shows the same slice at 200\,kyr.}
	\label{fig:slices}
\end{figure*}

\begin{figure*}
	\centering
	\begin{subfigure}{0.66\linewidth}
		\includegraphics[width=0.495\linewidth]{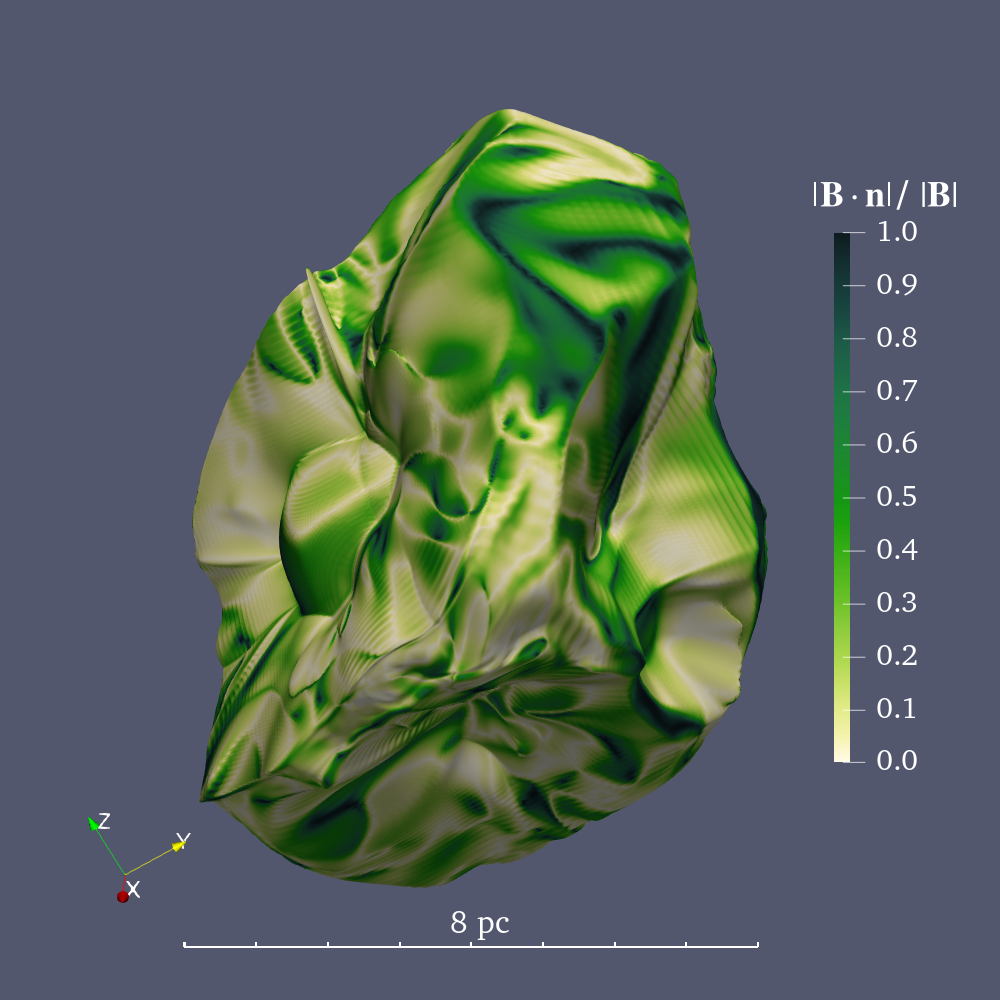}
		\includegraphics[width=0.495\linewidth]{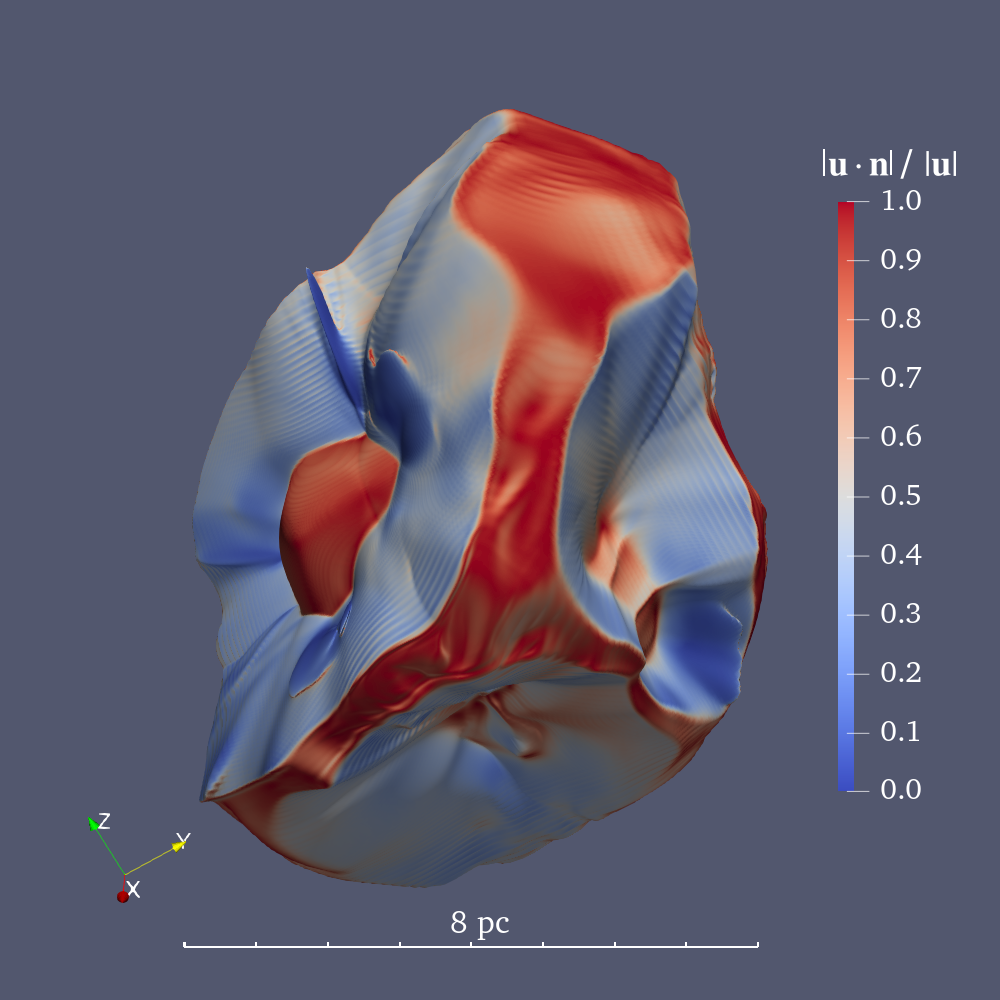}
		\caption{$M_\ur{S} = 3$, visualising the cluster-wind termination shock}
		\label{fig:ms3}
	\end{subfigure}
	\begin{subfigure}{0.33\linewidth}
		\includegraphics[width=0.991\linewidth]{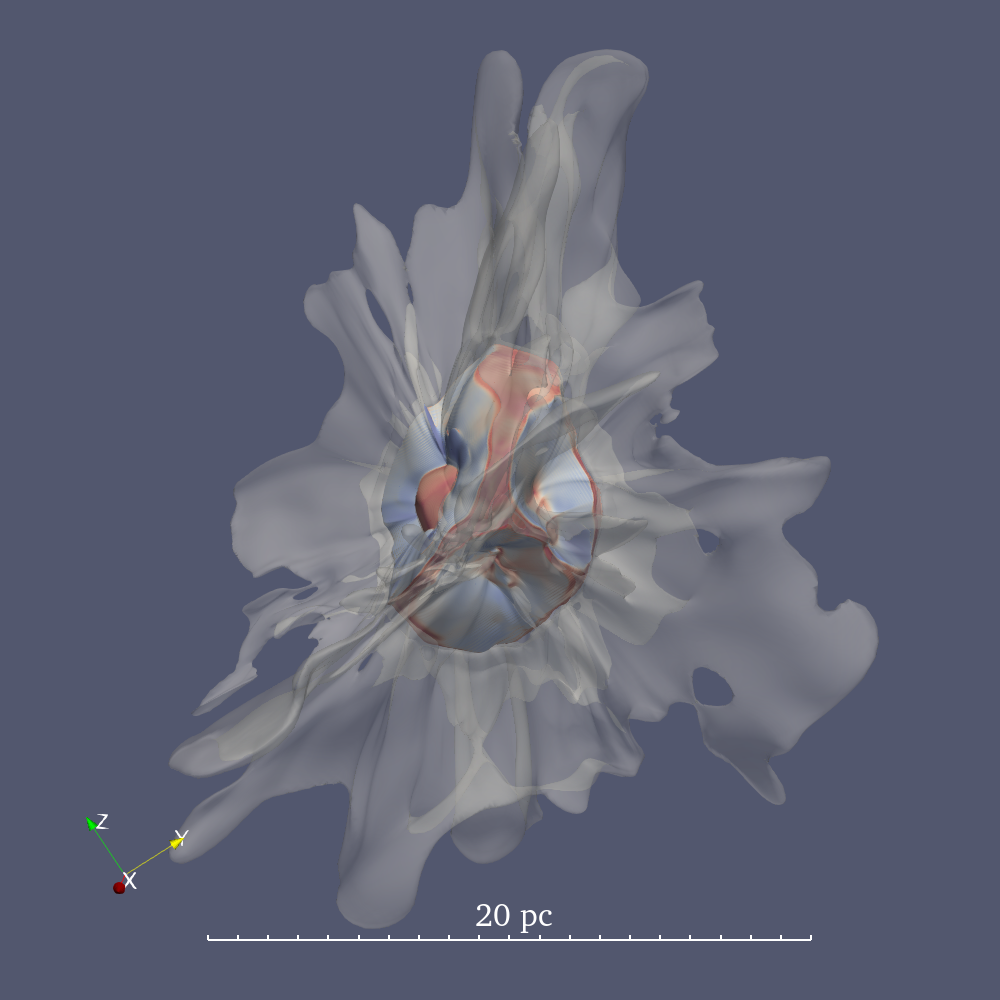}
		\caption{$M_\ur{S}=1$, visualising transonic sheets}
		\label{fig:ms1}
	\end{subfigure}
	
	\caption{Surfaces of constant sonic Mach number, visualising (a) the cluster-wind termination shock and (b) transonic sheets. The two panels in (a) show the orientation of the magnetic field, $\v{B}$ (left), and the flow velocity, $\v{u}$ (right), with respect to the Mach surface normal, $\v{n}$. In the latter plot, cones (blue, dominated by perpendicular flow) can be clearly distinguished from sheet base shocks (red, inbetween adjacent cones) and coupled stellar winds (red, in the centre of cones). In panel (b), the $M_\ur{S}=1$ surface is over-plotted onto the $M_\ur{S}=3$ surface surface in transparent white, highlighting that sheets extend outward from regions of parallel flow. The panels show cluster I at 390\,kyr.}
	\label{fig:ms} 
\end{figure*}


Figure~\ref{fig:rho-map} shows density slices in a region ${\pm}8\,$pc around the star cluster centre at $t\geq 200\,$kyr, ${\gtrsim}100\,$kyr after a supersonic cluster wind first emerges. We identify three main zones: the subsonic cluster core, the supersonic cluster wind, and the subsonic superbubble interior medium (``core'', ``cluster wind'', and ``superbubble'' hereafter). These three regions can be clearly distinguished in a Mach number plot (see Fig.~\ref{fig:ms-map}). Cluster wind and superbubble are separated by the cluster-wind termination shock (cluster WTS). We now discuss the formation of these regions and their evolution in detail.

\subsection{Formation and evolution of superbubble and cluster wind}
\label{sec:sb-ev}

Within 1--2\,kyr after initialisation, wind bubbles form around individual stars. The shells of these bubbles quickly encounter each other and break apart, forming a single, joint cluster-wind bubble. Yet individual wind termination shocks persist around individual stars (stellar WTSs). Material accumulates downstream of these shocks increasing pressure and density in the cluster core, which eventually launches a cluster wind. In its way out of the core, the wind flow is broken up and redirected by stellar WTSs, resulting in a complex, non-uniform flow morphology. At ${\sim}50\mbox{--}100\,$kyr, the cluster wind becomes supersonic and forms a cluster WTS beyond the core. The system enters a quasi-stationary state with stationary pressure in the core, stable flow geometry and a gradually expanding non-uniform cluster WTS. The size of the superbubble increases in accordance with 1D analytical theory for point-like injection (see Appendix~\ref{app:1d}). Since the system is quasi-stationary, we proceed to the WR phase at 200\,kyr, by increasing the mass-loss of the most massive star ten-fold. Up to the end of the simulation at 390\,kyr, five more stars enter the WR phase (see Sect.~\ref{sec:stars-pars} for details). Each new WR star modifies the flow and WTS geometry, as the distribution of injected wind power changes. Significant changes can occur within 10--20\,kyr (see Fig.~\ref{fig:rho-map}, panels b--c and e--f). The disruption of the existing structure visibly increases the dynamics of the downstream flow.  

\subsection{The central 5\,pc: subsonic core, stellar WTSs, cluster wind}

The star cluster and its immediate environment show stellar winds with Mach numbers ${>}100$ embedded in a bulk flow (see Fig.~\ref{fig:ms-map}). In the core, where the bulk flow is subsonic, stellar winds form WTSs, whose radii are determined by the balance of stellar wind ram pressure and thermal pressure in core bulk flow \citep[see][]{dyson72, weaver77}. Material processed through these shocks mixes efficiently before escaping the core as the cluster wind. These findings are in agreement with other work employing 3D simulations to study the core \citep{badmaev22, badmaev23, vieu24-core}. Stellar winds can also inject material directly into the cluster wind, in particular if they come from a star located at the edge of the cluster. The flow at the contact surface between stellar winds and cluster wind is highly oblique. Material passing this surface mixes less effectively with the bulk flow than it does in the core, which gives rise to density discontinuities extending in the radial direction (e.g.\ Fig.~\ref{fig:rho-map}, panels b--c). Some strong winds, such as the WR winds in the present work, fan out and extend all the way to the cluster WTS. We refer to these winds as ``coupled'' to the WTS. In contrast, stellar winds that are weaker or launched from deeper inside the cluster develop spherical or drop-like shapes. Both a larger separation between stars and a larger asymmetry in the injected wind power decrease the mixing of material into the bulk flow. The core density is on the order of a few particles per ${\rm cm}^{3}$ and the density in the wind can drop down to ${\sim}3\times10^{-2}\pccm$. The cluster WTS Mach number is ${>}5$ at 390\,kyr of simulation time.

\subsection{Transition to subsonic superbubble interior: cluster WTS and transonic sheets}
\label{sec:wts-sb}

On the time-frame of our simulations, the cluster wind terminates a few parsec away from the cluster core, producing a boundary layer which is discernible in Figs.~\ref{fig:rho-map} and \ref{fig:slices} by the jump in density, Mach number, and pressure. We identify this layer as the cluster WTS expected from analytical theory \citep[e.g.][]{dyson72, weaver77}. As briefly mentioned in Sect.~\ref{sec:sb-ev}, the cluster WTS is non-uniform. In particular the Mach number and radius are variable over the surface. The cluster WTS is bent inward by coupled stellar WTSs, giving rise to cone-like structures surrounding the coupled winds, as can be seen in Fig.~\ref{fig:ms3}. The bulk flow is funnelled outward in between the sections dominated by individual stellar winds and eventually becomes shocked and forms sheets (see Fig.~\ref{fig:ms-map} and Fig.~\ref{fig:ms1}), which can grow to scales of ${\sim}10\,$pc. In these sheets, the flow is compressed along the sheet normal by the pressure in the superbubble and re-accelerates to transonic speeds. In the first ${\sim}50\mbox{--}200\,$kyr, one-dimensional structures can be present in place of sheets, in particular for less dense clusters. These structures however do not persist to later times. Embedded within the sheets is a sequence of consecutive shocks, which weaken with increasing distance, the strongest shock being the cluster WTS at the sheet base (see Fig.~\ref{fig:divu-map}). This structure resembles ``shock diamonds'' and recollimation shocks observed in simulations of AGN jets \citep[e.g.][]{mizuno15}.
 
As the cluster WTS expands over time, individual winds can ``decouple'' and become fully embedded in the cluster wind. This can, though not always immediately, lead to a vanishing of the cone-like structure (see Fig.~\ref{fig:rho-map}, panels b--c and e--f). The timescale of the decoupling process depends strongly on the distribution of wind power in the cluster, the average distance between stars, and the position of the stars. While the cluster WTS tends to become more spherical over time, it is still visibly asymmetric at the end of our simulation at 390\,kyr, despite the rather compact model cluster (compare panels g and h in Fig.~\ref{fig:rho-map}), indicating that spherical WTSs may be the exception rather than the rule. The cluster I ($R_\ur{c}=0.6\,$pc) develops a stronger WTS earlier than cluster III ($R_\ur{c}=1\,$pc) and shows a larger fraction of decoupled WTS surface by the end of the simulation. Nevertheless, both models show the same qualitative WTS geometry. In panels g and h of Fig.~\ref{fig:rho-map}, the cluster WTS radius expected from 1D analytical theory is over plotted. Both panels show the same time, but different slices. The WTS is smaller than expected from 1D theory by ${\gtrsim}50\%$. This is because the sheets channel the winds' kinetic energy away from the core, which leads to a less efficient expansion of the cluster WTS.

\section{Magnetic field}
\label{sec:b}

\begin{figure*}
	\centering
	\includegraphics[width=\linewidth]{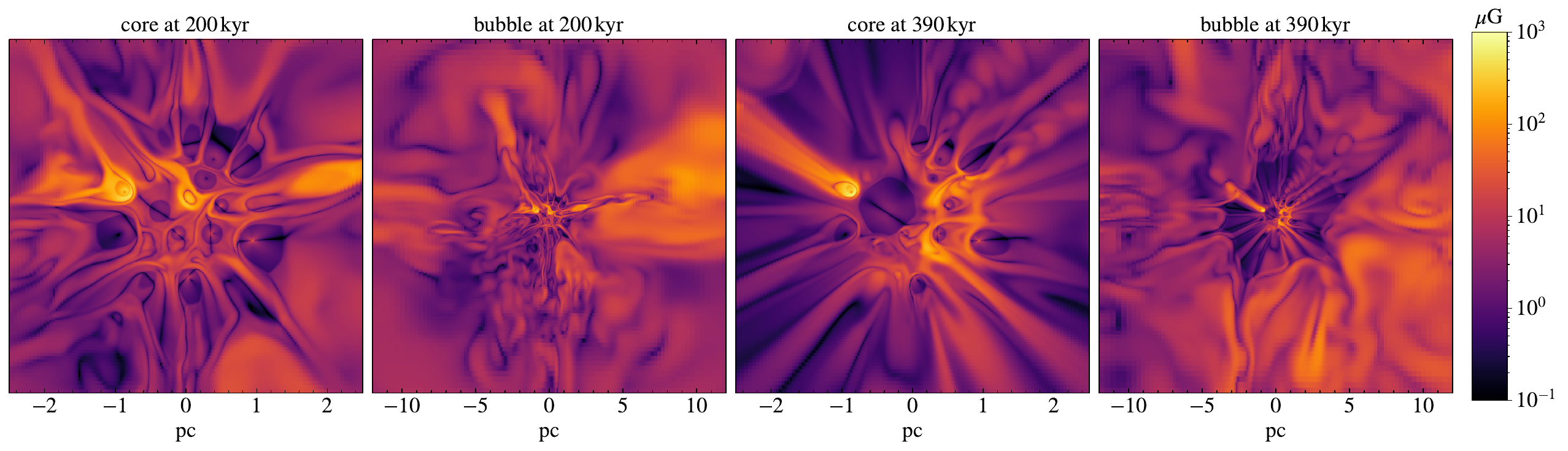}
	\caption{Slices of magnetic field magnitude, highlighting the non-uniformity of the field, which can partly be attributed to the five magnetic stars in the cluster. The magnetic field reaches 1\,mG in the cluster core and can fall to values of 0.1\,$\mu$G in the cluster wind. Large asymmetries persist into the superbubble, although they smooth out slowly over time. Two main observations are (1) field amplification behind the wind termination shock (e.g.\ right-most panel) and (2) mixing of flow from magnetic stars into the bulk medium (e.g.\ left-most panel). The figure shows data for cluster III sliced along the $z$-axis.}
	\label{fig:Bmap}
\end{figure*}

\begin{figure}
	\centering
	\includegraphics[width=0.8\linewidth]{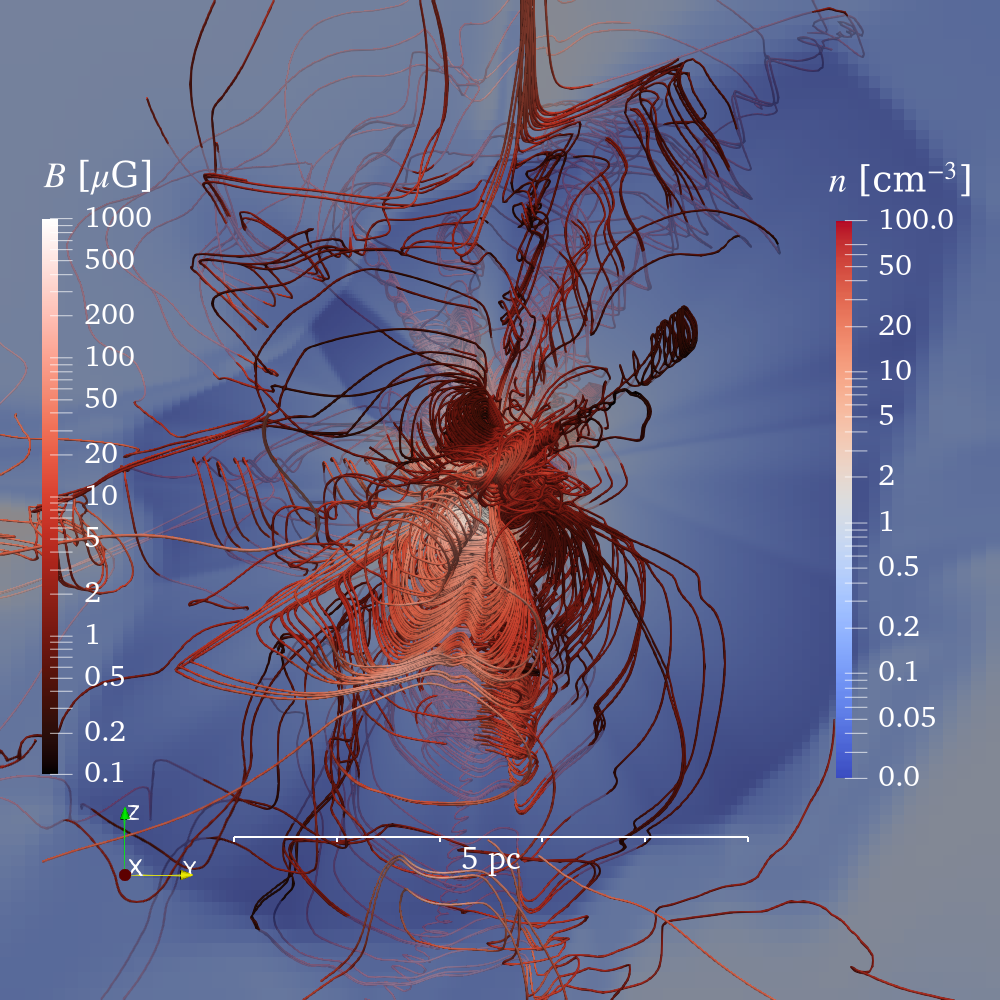}
	\includegraphics[width=0.8\linewidth]{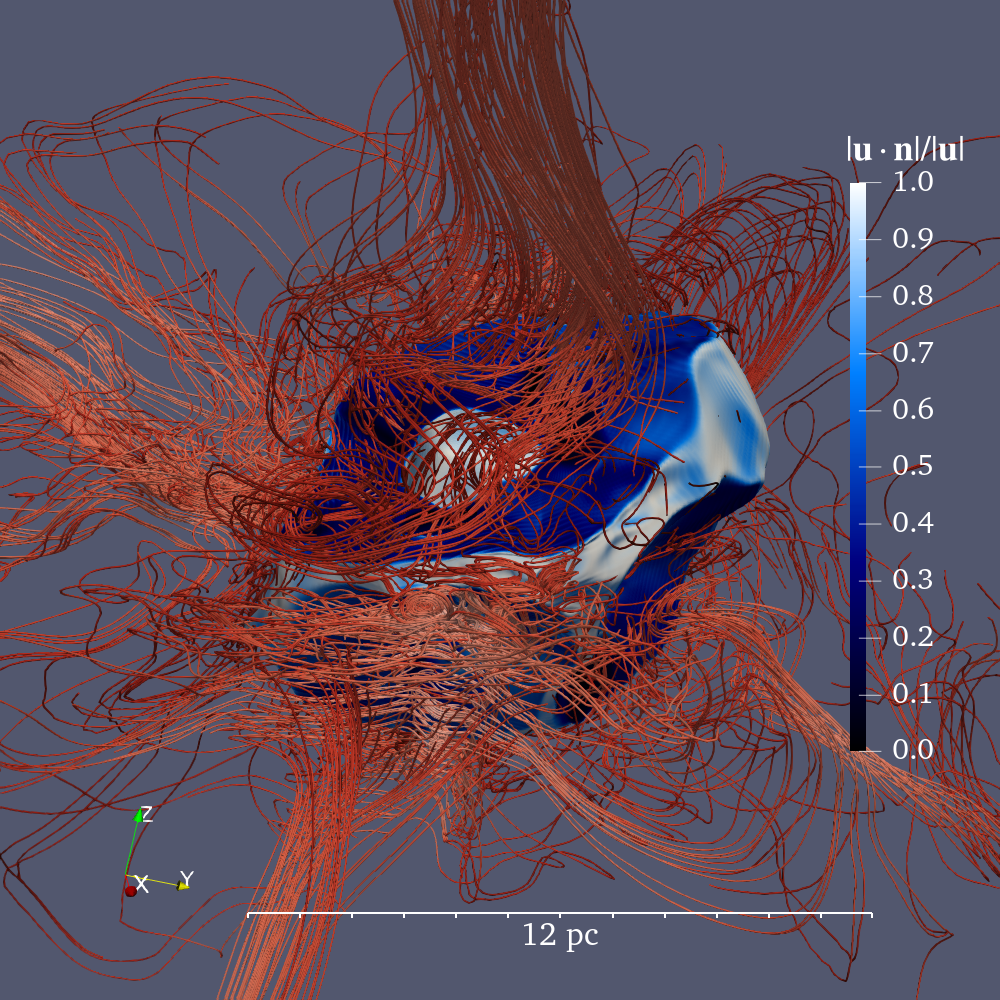}
	\caption{Streamlines of the magnetic field, coloured by its magnitude. In the upper panel, streamline seed points are placed within the radius of the star cluster. This highlights the tangled morphology arising from the interaction of Parker spiral stellar magnetic fields. The lower panel shows streamlines for a seed point radius comparable to the size of the cluster-wind termination shock (5\,pc), over-plotted onto the surface where the sonic Mach number equals three. The surface colour shows the orientation of the flow velocity, $\v{u}$, with respect to the surface normal, $\v{n}$. Note the quasi-radial streamline bundles, which are dragged outward by coherent flows. Such flows are present, for example, in transonic sheets (see Fig.~\ref{fig:ms1}). The figure shows cluster I at 390\,kyr.}
	\label{fig:strln} 
\end{figure}

\begin{figure*}
	\centering
	\includegraphics[width=0.85\linewidth]{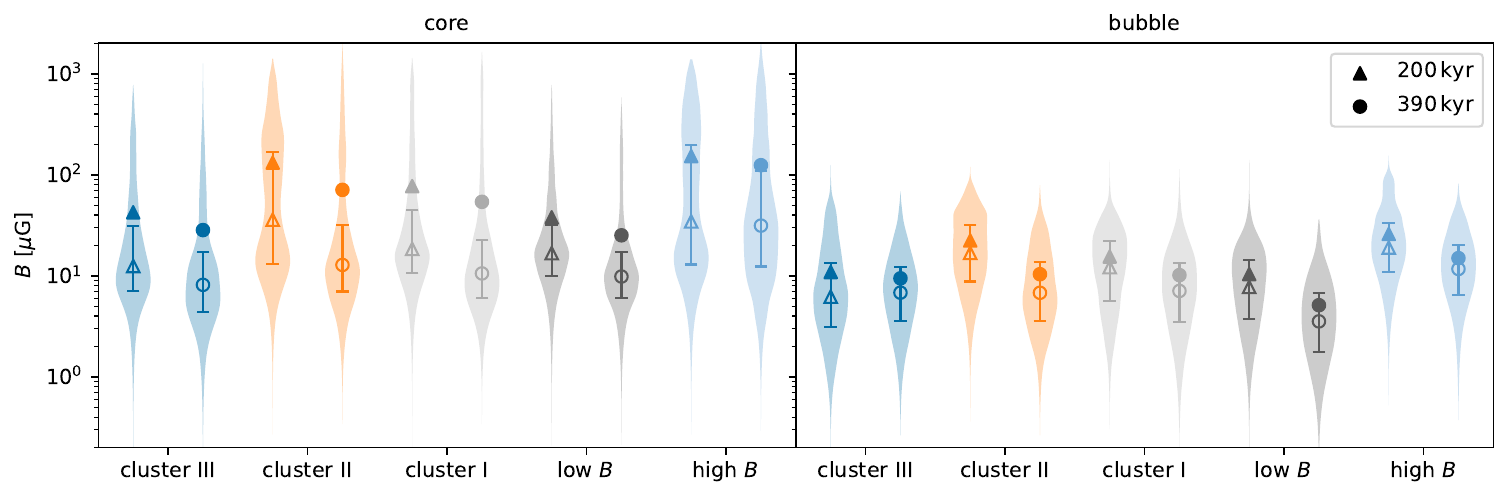}
	\caption{Magnetic field in the cluster core (left) and superbubble interior (right). Triangles mark distributions before the Wolf-Rayet onset at 200\,kyr and circles after the Wolf-Rayet onset at 390\,kyr. Colours indicate different simulation runs (see Tab.~\ref{tab:sims} for run parameters). Filled markers show the mean, empty markers the median. Whiskers indicate the 25th and 75th percentiles. Asymmetries due to local wind interactions are larger in the core, as evident by the broader distributions, but are reduced when the flow reaches the superbubble. Note that the distributions are shown in log space. }
	\label{fig:bav}
\end{figure*}

\begin{figure}
	\centering
	\includegraphics[width=\linewidth]{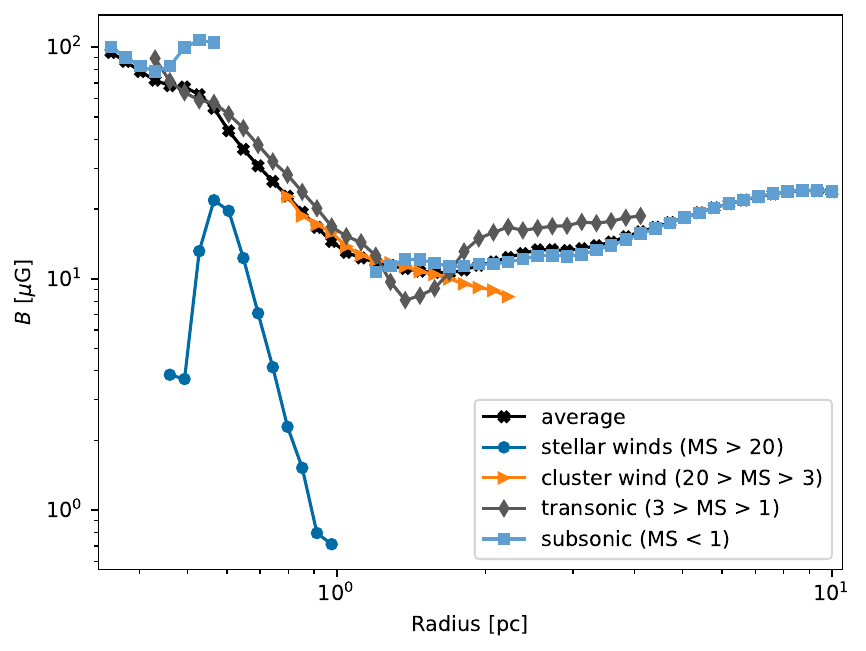}
	\includegraphics[width=\linewidth]{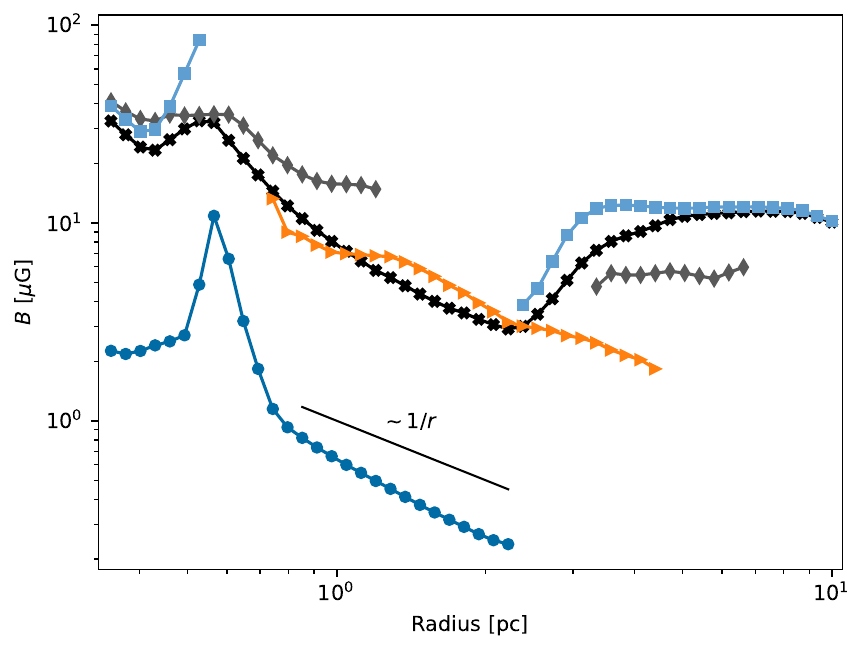}
	\caption{Radial profiles of the mean magnitude of the magnetic field, \Bm, in different regions for cluster II. The upper panel shows the profiles at 200\,kyr and the lower at 390\,kyr. The profiles were obtained by averaging over 10{,}000 randomly drawn line-outs. Points are only plotted where at least 10\% of line-outs pass through the indicated region. \Bm\ scales approximately as $r^{-1}$ in the supersonic cluster wind. The hump in the stellar winds profile close to $0.6\,$pc stems from magnetic stars and is traced by a hump in the subsonic core medium. This indicates that wind material from magnetic stars significantly increases the magnetisation of the bulk.}
	\label{fig:b-rad-log}
\end{figure}


We start this section with an overview in Sect.~\ref{sec:bqual}, illustrating the nature of the magnetic fields. This will be followed-up by deriving average properties in Sect.~\ref{sec:bav}--\ref{sec:fldiff}.

\subsection{Qualitative overview}
\label{sec:bqual}

Stars are initialised with randomly oriented Parker spiral magnetic fields. The global magnetic field is determined by how these fields interact, in interplay with the flow. The flow is not magnetically dominated (Alfv\'enic Mach number $M_\ur{A} \gg 1$, see Fig.~\ref{fig:ma-map}). Therefore the flow determines the dynamics of the magnetic field. In ideal MHD, magnetic fieldlines are frozen into the flow. Parker spirals get deformed (e.g.\ stretched or twisted) as fieldlines are carried in the intricate flow pattern described in Sect.~\ref{sec:evo}. This results in a non-trivial magnetic field morphology, which strongly depends on local wind interactions.

Figure~\ref{fig:Bmap} shows the magnetic-field magnitude for cluster III in different regions and at simulation times 200\,kyr and 390\,kyr. The magnetic field is non-uniform, reaching values of 1\,mG close to magnetic stars in the core and dropping to ${<}1\,\mu$G in parts of the cluster wind and superbubble. In the superbubble, large patches with comparatively high $B$ can be found where the flow is dominated by winds from magnetic stars. The magnetic field in the core ambient medium is fed by magnetic stars and higher on average than the field in the winds of non-magnetic stars. Asymmetries tend to be smoothed out in the superbubble and $B$ varies on larger spatial scales. We observe magnetic field amplification in the immediate downstream of the cluster WTS. 

Figure~\ref{fig:strln} shows magnetic field streamlines. Close to the cluster (upper panel), the field is highly tangled and locally dominated by Parker spirals from individual stars. The lower panel of Fig.~\ref{fig:strln} shows that streamlines trace the downstream WTS surface, which is expected from amplification of the perpendicular field component at the WTS. Transonic sheets harbour coherent streamline bundles, extending outward in the radial direction. Coherent streamline bundles are not only found in transonic sheets, but also in coherent subsonic flows, for example, those extending beyond coupled WR winds. Beyond the transonic sheets, the field becomes increasingly disorganised, but all streamlines eventually connect back to the core.

\subsection{Average magnitude and distribution}
\label{sec:bav}

A critical question for particle acceleration and transport is what magnitude $B$ takes in the different regions. We obtain distributions and volume weighted mean and median values of $B$ in the subsonic regions (i.e., the core and superbubble). The volume weighting is necessitated by the non-uniform grid we employ. The values for different regions are computed by filtering based on Mach number and radius.

Figure~\ref{fig:bav} shows the distributions of $B$ in the core and superbubble for all simulation runs. The mean, median, 25th, and 75th percentiles are marked. All distributions show a tail at large values, in other words, the mean consistently lies above the median. The most skewing is seen in the core, where the magnetic fields are kept fixed inside the stellar boundaries. The least skewing is seen in the bubble, indicating that averaging effects take place. The mean of $B$ lies between $30\mbox{--}200\,\mu$G in the core and $5\mbox{--}25\,\mu$G in the superbubble. The median values of $B$ are $8\mbox{--}35\,\mu$G in the core and $4\mbox{--}20\,\mu$G in the superbubble. Note that the flow in both the core and the superbubble is turbulent. By construction, we can only investigate the field modes above the grid resolution. The grid resolution is non-uniform, due to the stretched grid we employ. This leads to a suppression of $B$ in small-scale modes in the superbubble. 

In the core, the distribution of $B$ shows noticeable variations between cluster I--III, despite their identical magnetic star content. In contrast to cluster I, cluster II shows clear bimodality at 200\,kyr, even though both clusters have identical radii. This indicates that the distribution of $B$ is influenced significantly by local wind interactions. The most powerful magnetic star is located close to the cluster centre for cluster II, while it is closer to the outskirts in cluster I (see Fig.~\ref{fig:cluster3d} in App.~\ref{app:tab-stars}). The larger $B$ in cluster II could therefore be a consequence of increased mixing of magnetic wind material into the core bulk medium. In the superbubble, cluster II also shows a distribution shifted to higher $B$ at 200\,kyr, but shows a distribution equivalent to those of clusters I and III at 390\,kyr. This suggests that the distribution of $B$ in the superbubble is not influenced by the efficiency of mixing processes in the core after a few 100\,kyr.  

Since the superbubble is not a stationary system, median $B$ is not stationary. Before the WR onset, the decrease in the $B$ median values is on average consistent with that expected from the increase in superbubble volume, assuming that magnetic energy is diluted by the expansion of the bubble. After the WR onset, the observed decrease exceeds the one expected from superbubble expansion alone by a factor of about two. All WR winds originate from non-magnetic stars. Therefore, the average wind magnetisation goes down in the WR phase, which can contribute to decrease $B$.  

\subsection{Radial profile}
\label{sec:rprofile}

In Sect.~\ref{sec:evo}, several distinct domains were identified, including the subsonic core bulk flow, stellar winds, the cluster wind, transonic sheets, and the subsonic superbubble medium. We obtain radial profiles of the mean magnitude of the magnetic field, \Bm, in these regions by filtering on the Mach number. The radial profiles are obtained by averaging over a large number of line-outs ($10{,}000$). Note that the variation between single line-outs is large (see the large variations in Fig.~\ref{fig:Bmap}). 

Figure~\ref{fig:b-rad-log} shows the radial profile of \Bm\ for cluster II before the WR phase at 200\,kyr and 390\,kyr. Radial profiles for the remaining simulation runs can be found in appendix~\ref{app:b-suppl}. At 390\,kyr, the radial domain can be broadly subdivided into a region dominated by subsonic core medium, a region dominated by supersonic wind, and a region dominated by subsonic superbubble medium. The transition between the first two is at ${\sim}0.8\,$pc, and between the latter two at $2\mbox{--}3\,$pc for a cluster radius of 0.6\,pc. $B$ decreases slighly steeper than $r^{-1}$ in the region dominated by supersonic cluster wind. Outside the wind dominated region, it approaches a constant value. 

Stellar winds ($M_\ur{S}>20$) show values below the overall average by a factor $20\mbox{--}30$ and a peak at ${\sim}0.6\,$pc, which is traced by a peak in the subsonic profile. The peak in the stellar wind profile stems from the distribution of magnetic stars. The sharp decline following the peak is an averaging effect: wind from magnetic stars skew the average upward, but contribute only in a narrow range of radius. Only WR winds, which are all non-magnetic, extend out to larger radii. The fact that the peak is traced by the subsonic profile indicates that magnetic stars contribute substantially to the magnetisation of the core bulk medium and cluster wind.  

Transonic sheets have \Bm\ close to the overall average at 200\,kyr and \Bm\ below the average by about a factor of two at 390\,kyr. Before the WR phase, all profiles are shifted to higher \Bm\ values, in line with what was discussed for the $B$ distributions in the preceding section. In addition, a larger fraction of the cluster wind is still in the transonic regime ($3>M_\ur{S}>1$, see Fig.~\ref{fig:b-rad-log}).

\subsection{Streamline diffusivity}
\label{sec:fldiff}

\begin{figure*}
	\centering
	\includegraphics[width=0.49\linewidth]{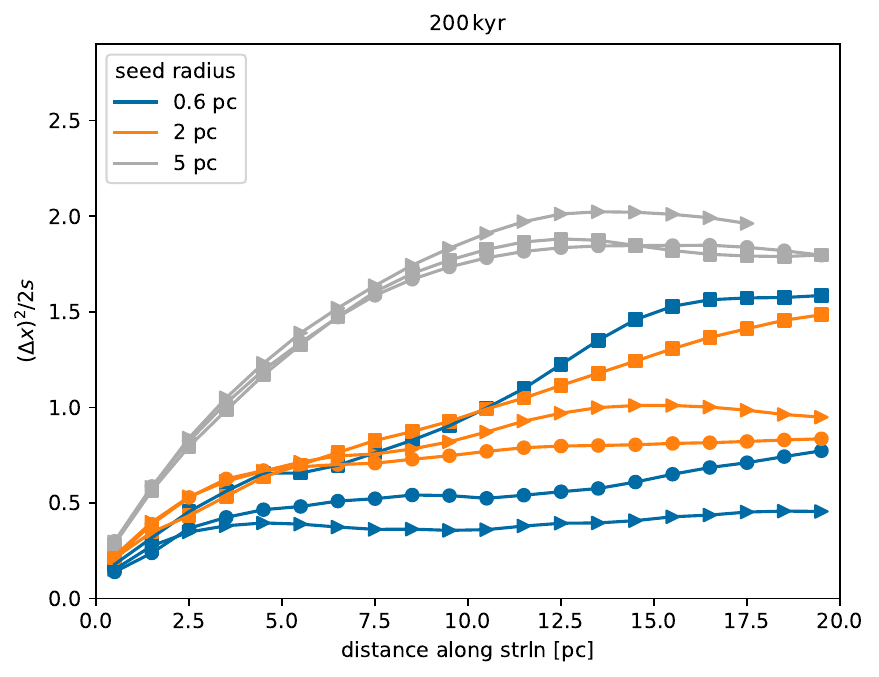}
	\includegraphics[width=0.49\linewidth]{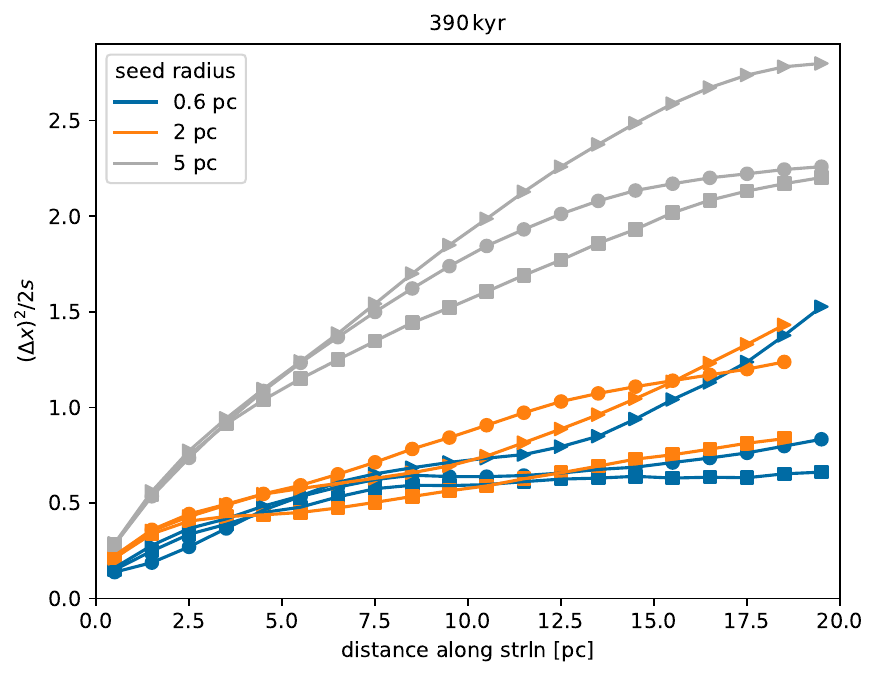}
	\caption{Streamline diffusivity for seed points placed within spheres of the indicated radii at 200\,kyr (left) and 390\,kyr (right). Triangles, circles, and squares indicate clusters I, II, and III, respectively. The near linear increase for the 5\,pc seed cloud radius is due to quasi-radial field-line bundles embedded in transonic sheets (cf.~Fig.~\ref{fig:strln}, lower panel) and the levelling-off is due to fieldlines travelling close to the superbubble edge looping back to the core. The variation between clusters I and II, which only differ in the spatial distribution of stars, illustrates the importance of local wind interactions.}
	\label{fig:fldiff}
\end{figure*}


As discussed in Sect.~\ref{sec:bqual}, the magnetic field shows both tangled sections and coherent structures, such as spirals, loops, and field-line bundles. No simple description for this highly intricate morphology can be provided. Here, we investigate streamline diffusivity to provide a measure for the average behaviour of the field. It also has implications for particle transport, a discussion of which will follow in Sect.~\ref{sec:disc}. The streamline diffusivity is defined here as $D = \langle \Delta x^2\rangle /2s$, where $\Delta x$ is the spatial separation of two points on a streamline and $s$ the distance along the streamline. Note that, in general treatment, $D$ is a tensor often expressed in terms of components parallel and orthogonal to a local mean field. Since the mean field is not well defined here, we resort to the above scalar definition. We integrate streamlines using \texttt{Paraview}, distributing 1000 streamline starting-points (seeds) uniformly inside a spherical region centred on the star cluster. We investigate three different seed cloud radii (0.6\,pc, 2\,pc, and 5\,pc), corresponding to the size of the cluster core, cluster wind, and cluster WTS regions.   

Figure~\ref{fig:fldiff} shows the diffusivity at 200\, and 390\,kyr. We compare clusters I--III. Over all runs and timesteps, the diffusivity is in the range of $D = \langle \Delta x^2\rangle/(2s)\sim0.5\mbox{--}2\,$pc. Thus the linear distance is
\begin{equation}
\Delta x = \left( \frac{D}{0.5\,\ur{pc}} \right)^{0.5} \left( \frac{s}{1\,\ur{pc}} \right)^{0.5} \, \ur{pc} \, ,
\end{equation}   
from which it follows that on distances larger than a parsec, field-line diffusion can play an important role in particle transport. The diffusivity is approximately constant for seeds placed in the core and wind, with some deviation for cluster III at 200\,kyr and cluster I at 390\,kyr. For a seed cloud radius of 5\,pc, the diffusivity is clearly increasing with $s$, although at 200\,kyr, it levels off starting at around $10\,$pc. This is due to the fact that streamlines reach the edge of the superbubble and loop back to the core. Since the superbubble is larger at 390\,kyr, this levelling-off starts at a larger $s$. The close-to-linear increase prior to this comes from coherent field-line bundles extending outward in transonic sheets. Linear behaviour is expected if streamlines are straight lines. Between 200\,kyr and 390\,kyr, diffusivity has decreased for cluster III ($R_\ur{c}=1\,$pc), while the opposite is true for cluster I ($R_\ur{c}=0.6\,$pc). Jointly with the variation between cluster I and II, which only differ by the spatial distribution of stars, this indicates the importance of local wind interactions for the diffusivity. Nevertheless, the difference does not exceed a factor of a few.

\section{Implications for particle acceleration and transport}
\label{sec:disc}

\begin{figure*}
	\includegraphics[height=0.32\linewidth]{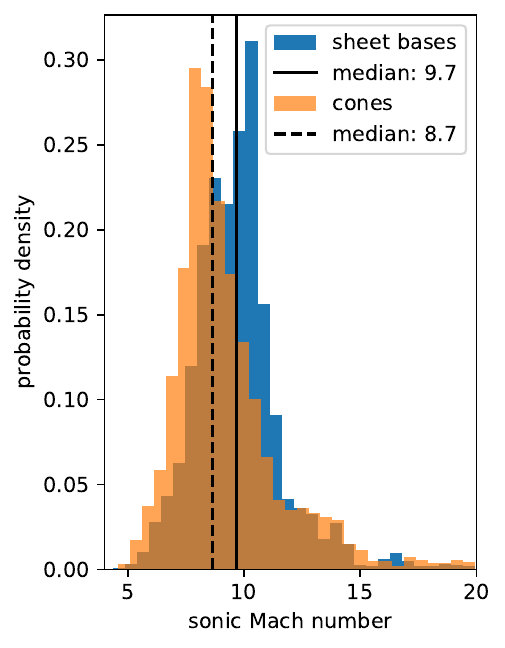}
	\includegraphics[height=0.32\linewidth]{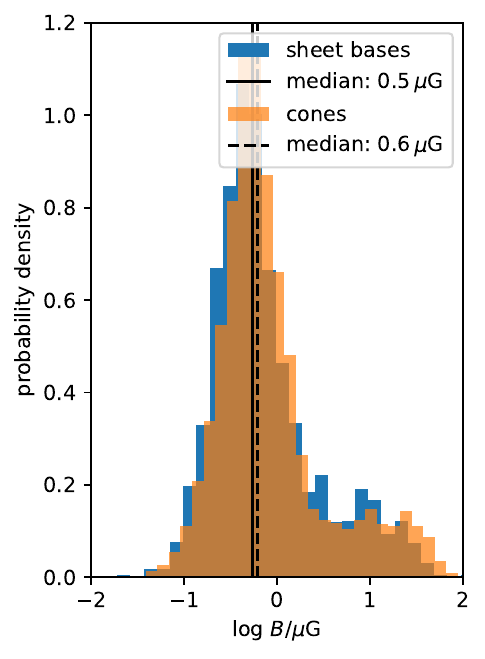}
	\includegraphics[height=0.32\linewidth]{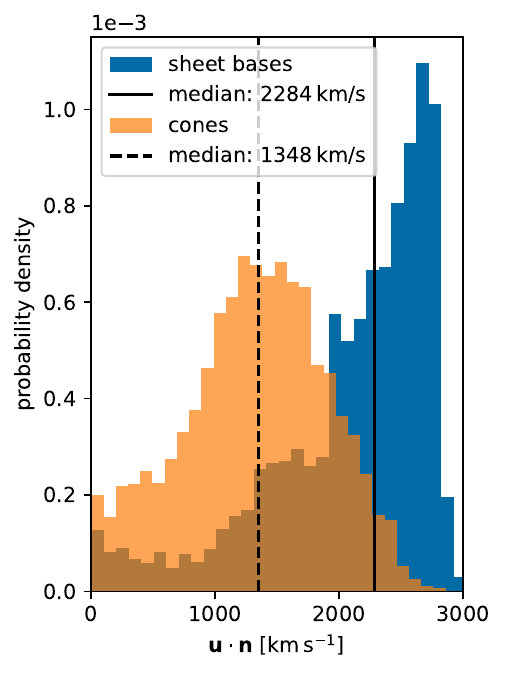}
	\includegraphics[height=0.32\linewidth]{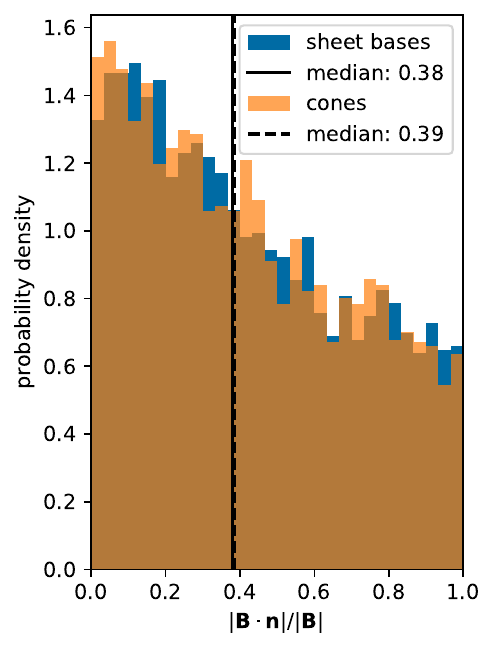}
	\caption{Properties in the immediate upstream of sheet base shocks and wind cones at 390\,kyr of simulation time. The histograms were compiled from values along 10{,}000 randomly drawn line-outs. The distribution of values at sheet bases and cones is similar for most parameters, except the velocity across the shock, $\v{u}\cdot\v{n}$ (panel 3). For further description, see the text.}
	\label{fig:upstr-vals}
\end{figure*}

\begin{figure}
	\centering
	\includegraphics[width=\linewidth]{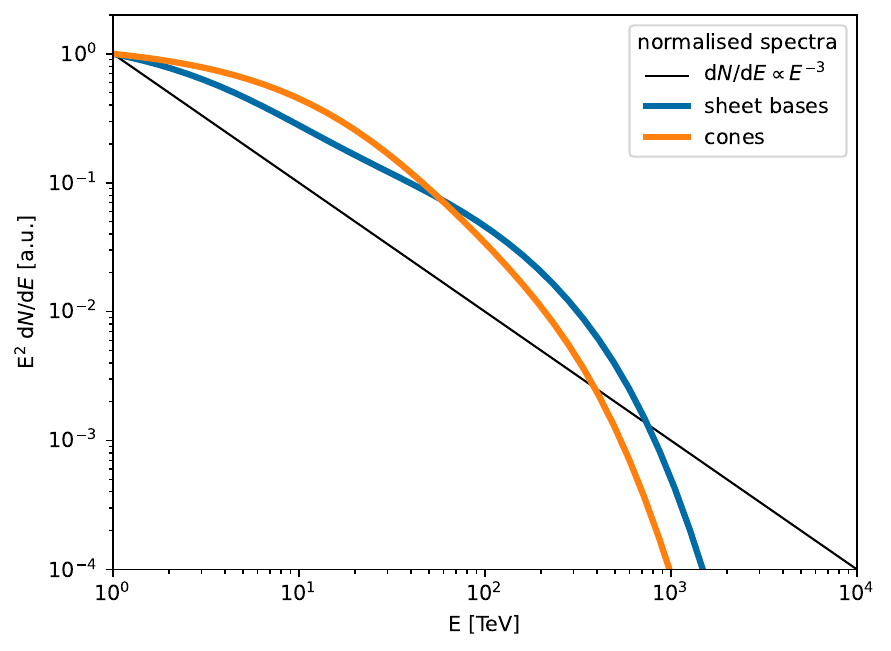}
	\caption{Toy model for the spectra of particles accelerated at sheet base and cone shocks for cluster I. The maximum cut-off energy is close to 500\,TeV. For further description, see the text. }
	\label{fig:spec}
\end{figure}


The results discussed in Sect.~\ref{sec:evo} and \ref{sec:b} present favourable conditions for particle acceleration at stellar and cluster WTSs within the framework of diffusive shock acceleration (DSA, \citealt{axford77, krymskii77, bell78a, blandford78}). In the discussion to follow, we assume that DSA operates at maximum efficiency. In such cases, the maximum particle energy is predicted to be close to the Hillas limit: $E_\ur{max} = ZeBru/c$, where $q=Ze$ is the charge of the accelerated particle, $u$ the flow speed, $B$ the magnetic field and $r$ the size of the accelerator \citep[see][]{LagageCesarsky,hillas84}. We discuss acceleration at individual stellar WTSs and the cluster WTS in turn in Sect.~\ref{sec:acc-core} and \ref{sec:cl-acc}. In Sect.~\ref{sec:b-concl}, sub-grid effects and their potential impacts are examined. We then discuss a toy model for the spectrum in Sect.~\ref{sec:spec-model} and finally turn to particle transport in Sect.~\ref{sec:transp}.

\subsection{Acceleration at stellar WTSs in the core}
\label{sec:acc-core}

Stellar WTSs in the core have radii of about $0.03\mbox{--}0.3\,$pc and are strong shocks, $M_\ur{S}\gg 10$ (see Fig.~\ref{fig:ms-map}). \citet{vieu24-core} discussed collective effects in the core, and concluded that they do not increase the maximum energy, nor have a significant impact on the spectrum in all considered scenarios. This is mainly because the volume filling factor of the powerful winds in the core is rather small, even for compact clusters, which does not allow particles to interact with multiple shocks within their (short) advection time out of the core. Stellar WTSs in a cluster can therefore be viewed as independent, individual accelerators. The magnetic field in the upstream of individual stellar WTSs is a Parker spiral by initialisation (see Eq.~\ref{eq:parker1} and \ref{eq:parker2}). The azimuthal component, $B_\varphi$, dominates far from the stellar surface, $r\gg R_\star$. Since, in the Hillas limit, $E_\ur{max}\propto Br$, but $B_\varphi \propto r^{-1}$ for a Parker spiral, $E_\ur{max}$ can be expressed in terms of the surface field, $B_0$. For fiducial parameters of a magnetic star this gives
\begin{equation}
E_\ur{max} < 21\times Z \left( \frac{B_0}{1\,\ur{kG}} \right) \left( \frac{R_\star}{10\,\ur{R}_\odot} \right) \left( \frac{0.01}{u_\ur{rot}/u_\ur{w}} \right)  \left( \frac{u_\ur{w}}{3000\,\kms} \right)  \,\ur{TeV} \, .
\label{eq:emax-core}
\end{equation}
where the ratio $u_\ur{rot}/u_\ur{w}$ enters through $B_\varphi$. Standard stars ($B_0=10\,$G) in our model cluster have $E_\ur{max} < 2.5\mbox{--}5.4\,$TeV and magnetic stars ($B_0=1\,$kG) have $E_\ur{max} < 240\mbox{--}450\,$TeV. The latter constraint should be taken as optimistic, since we set $u_\ur{rot}=300\,\kms$ for all stars and the magnetic field possibly scales down steeply close to the stellar surface (see Sect.~\ref{sec:b-setup}). Even in the exceptional case of a high surface field and fast rotation, $E_\ur{max}$ is limited by the fact that DSA requires an Alfv\'enic Mach number $M_\ur{A}\gg 1$. This places an absolute limit on the maximum energy. With $M_\ur{A} = u_\ur{w}/u_\ur{A}$, where $u_\ur{A} = B/\sqrt{4\pi\rho}$, and the density falling off as $\rho = \dot{M}( 4\pi r^2 u_\ur{w} )^{-1}$, we obtain
\begin{multline}
	E_\ur{max} \ll 220 \times Z \left( \frac{\dot{M}}{10^{-6}\,\Msyr} \right)^{0.5} \left( \frac{u_\ur{w}}{3000\,\kms} \right)^{1.5} \\ \times \left( \frac{M_\ur{A, min}}{3} \right)^{-1} \,\ur{TeV} \, ,
	\label{eq:emax-abs}
\end{multline}
where we impose a minimum Alfv\'enic Mach number, $M_\ur{A, min}$, of 3. For stars in our model cluster, the above limit is 100--330\,TeV in the main sequence phase and 1.7\,PeV for the most powerful WR star.

In conclusion, the WTS of a typical massive star is not expected to accelerate particles to energies beyond ${\sim}10$\,TeV. Higher energies can only be reached by magnetic stars with $B\gtrsim 100\,$G, which make up ${\lesssim}10\%$ of O stars \citep[see][]{grunhut17}. Even a fast rotating strongly magnetic star, a case which is both observationally and theoretically expected to be exceptionally rare, can reach at most ${\sim}1\,$PeV. An additional aspect of particle acceleration at stellar WTSs is that the shock is expected to be highly oblique in $\v{B}$, since the $B_\varphi$ component dominates at the WTS. The implications of magnetic-field obliquity on particle injection efficiency and maximum energy are not fully conclusive \citep[e.g.][]{Bell2011,Xu2020,Kumar2021}.  

\subsection{Acceleration at the cluster WTS}
\label{sec:cl-acc}

The cluster wind terminates at a non-uniform shock surface beyond the core, which is made up of coupled stellar WTSs, sheet base shocks, and cones (see Sect.~\ref{sec:wts-sb}). The discussion in the preceding section also applies for coupled stellar WTSs. Their surface is larger than that of stellar WTSs in the core. In $E_\ur{max}$, however, this is compensated by the fact that $B$ is scaled down to a lower value ($B\propto r^{-1}$ approximately, see Fig.~\ref{fig:b-rad-log}). 

The discussion to follow focuses on the collective part of the cluster WTS, which has $M_\ur{S}\sim 10$ (see Fig.~\ref{fig:ms-map}). We extract properties at sheet bases and cones along 10{,}000 randomly drawn radial line-outs. The immediate shock upstream is taken to be the location where $M_\ur{S}$ is maximal. Sheet bases and cones differ in the obliquity of the flow (see Fig.~\ref{fig:ms3}). We take the field of normal vectors of the cluster WTS surface to be $\v{n} = -\nabla M_\ur{S}/\vert \nabla M_\ur{S} \vert$. A threshold of $\v{u}\cdot\v{n}/|\v{u}|>0.7$ was found to reliably filter for line-outs passing through sheet bases. The inverse filter identifies cones. Coupled stellar winds are removed by imposing $M_\ur{S}<20$. The high obliquity of the flow in cones calls for a verification that they can be considered shocks as opposed to shear layers. The ratio $\mathcal{R}=\nabla \cdot \v{u}/\nabla \times \v{u}$ quantifies the relative contribution of compression and shear. Less than 1\% of line-outs have $|\mathcal{R}| < 1$ in the downstream of the cones for all simulation runs. In addition, the divergence of $\v{u}$ shows similar values across all parts of the cluster WTS (see Fig.~\ref{fig:divu-map}), with only slightly lower values in the cones. 

Figure~\ref{fig:upstr-vals} shows the immediate upstream distribution of a selection of parameters for cluster I at 390\,kyr. The sonic Mach number is characterised by a symmetric distribution with median values of about $10$ in sheet bases and $9$ in cones and $M_\ur{S} \gtrsim 5$ everywhere, indicating strong shocks with a compression ratio $q>3.5$. The median value of $B$ is 0.5\,$\mu$G in sheet bases and 0.6\,$\mu$G in cones. The distribution of $\log_{10} B$ shows a hint of bi-modality, with the second mode peaking at about $10$\,$\mu$G. An examination of the distribution of magnetic field values on the cluster WTS surface reveals that values above about $3\,\mu$G almost exclusively appear in the flow-paths of magnetic stars. In these regions, upstream values can reach averages of about $20\,\mu$G over 1\,pc. Note that cluster II shows a significantly lower maximum $B$ in the upstream of sheet base shocks (see Fig.~\ref{fig:bupstr-other-sims}). The fact that the most powerful magnetic star is positioned closer to the centre of the cluster seems to increase the level of mixing, leading to a higher average $B$ (see Fig.~\ref{fig:bav}), but fewer high $B$ outliers. The third panel in Fig.~\ref{fig:upstr-vals} shows the flow speed across the shock, $\v{u}\cdot\v{n}$. The speed across the shocks is significantly lower in the cones than it is in the sheet bases, where the modes of the distributions are at about $1400\,\kms$ and $2600\,\kms$, respectively. The latter value is close to the median magnitude of the velocity, $u$, which is about $2800\,\kms$, with a width of ${\pm}50\,\kms$ at half maximum. 
Both distributions are rather broad and show a tail at low values. These tails are due to contamination of the sample with downstream values and the rotation of $\nabla M_\ur{S}$ in the numerically smoothed shock. We use $\nabla M_\ur{S}$ to determine the shock normal vector. The right-most panel shows that the orientation of $\v{B}$ in the shock upstream is highly variable. The median value lies close to 0.4 for both cones and sheet bases. 

Thus far, we have demonstrated that both cones and sheet bases are strong shocks with a median upstream magnetic field of $0.5\,\mu$G, a broad range of obliquities in $\v{B}$, and lower flow speed normal to the surface in the case of the cones, owing to the obliquity of the flow. Despite these topological subtleties, DSA is expected to take place at both cones and shear bases. The acceleration properties however depend on the local parameters of the flow and magnetic field. In the following, the maximum particle energy is estimated case by case. The width of sheet base shocks is $d\sim0.5\mbox{--}3\,$pc, with a typical value of 1\,pc. The median value of 0.5\,$\mu$G gives
\begin{equation}
E_\ur{max}^\ur{sheetbase} = 4.0\times Z \left( \frac{B}{0.5\,\mu\ur{G}} \right) \left( \frac{d}{1\,\ur{pc}} \right) \left( \frac{u}{2600\,\kms} \right)  \,\ur{TeV} \, .
\label{eq:max-shb}
\end{equation}
The cones reach a size comparable to the overall extent of the cluster WTS, which is about $5\,$pc. The decreased normal flow velocity gives a comparable limit,
\begin{equation}
E_\ur{max}^\ur{cones} = 11\times Z \left( \frac{B}{0.5\,\mu\ur{G}} \right) \left( \frac{d}{5\,\ur{pc}} \right) \left( \frac{u}{1400\,\kms} \right)  \,\ur{TeV} \, .
\label{eq:max-con}
\end{equation}  
Since the distribution of $B$ is rather broad, $E_\ur{max}$ can reach higher values in some regions. In the isolated regions where the field is about $20\,\mu$G over 1\,pc, $E_\ur{max}$ is ${\lesssim}170$\,TeV. Note that the constraints on $E_\ur{max}$ are not expected to change considerably when the WTS evolves over a time longer than that covered by the simulation. In the limit of a fully spherical cluster WTS and radial cluster wind, $u$ across the shock is given by the median magnitude, $2800\,\kms$, and $d$ is the the size of the cluster WTS. This increases $E_\ur{max}$ by a factor two compared to the limit in the cones. The expansion of the cluster WTS over time does not per se increase $E_\ur{max}$, since $B$ drops in the cluster wind approximately as $r^{-1}$ (see Fig.~\ref{fig:b-rad-log}). As a consequence, in $E_\ur{max}$, the increase in size of the accelerator cancels with the decrease in $B$. 

In this and the preceding section, we have discussed $E_\ur{max}$ at stellar WTSs and the cluster WTS, informed by our simulations. We find that $E_\ur{max}$ does not exceed a few 10s of TeV in typical scenarios, even when extrapolating our results to a quasi-spherical cluster WTSs grown to 10s of pc over a timescale of Myr. Regions of the cluster WTS with higher than average $B$, for example in the flow-paths of magnetic stars, reach at most a few 100s of TeV. Individual magnetic stars are unlikely to accelerate particles to these energies (see Eq.~\ref{eq:emax-core}) and in fact have a theoretical upper bound on $E_\ur{max}$ close to 200\,TeV for typical parameters. Note that we purposefully left out a discussion of supernovae (SNe). SNe expanding in the collective wind of a compact cluster can potentially reach higher energies than the cluster WTS \citep{vieu23}. A dedicated analysis of the evolution of SNe in a star-cluster environment is necessary to assess the properties of acceleration at their shocks.

\subsection{Sub-grid effects}
\label{sec:b-concl}

In this section we explore two mechanisms to generate magnetic field on sub-grid scales, which could ultimately facilitate acceleration to higher energies than estimated in the two preceding sections: sub-grid dynamos in Sect.~\ref{sec:dyno} and magnetic field excitation by streaming instabilities in Sect.~\ref{sec:streaming}. 

\subsubsection{Dynamos}
\label{sec:dyno}

The maximum energy at the cluster WTS is most decisively set by $B$. We find a typical upstream value of ${\sim}0.5\,\mu$G. This value is set by the value of $B$ in the core and its scaling in the supersonic cluster wind. A higher $E_\ur{max}$ at the cluster WTS could therefore be facilitated by a higher $B$ in the core. Even considering that limitations inherent to the numerical approach could affect our measured $B$, the scaling is always expected to be $r^{-1}$ or steeper (see Sect.~\ref{sec:rprofile} and \ref{sec:acc-core}). In an optimistic scenario ($d=5\,$pc and $u=2800\kms$), an upstream value of 30$\,\mu$G could facilitate acceleration to PeV energies. This would require a bulk magnetisation in the core of at least 150$\,\mu$G, assuming the field scales as $r^{-1}$ in the cluster wind and disregarding potential amplification by streaming instabilities. In our simulation, the magnetisation of the bulk flow in the core is modest, with typical values of 8--35\,$\mu$G. In principle, dynamo effects below the grid scale ($0.008\,$pc in the core) could increase the value of $B$. However, dynamo effects would have to amplify $B$ by a factor 5--20 over the simulation value to reach the 150$\,\mu$G. In conclusion, small scale (${<}0.01\,$pc) dynamo effects would have to be by far the dominant effect supplying the magnetic field to facilitate particle acceleration to PeV energies at the cluster WTS.

\subsubsection{Self-excited magnetic fields}
\label{sec:streaming}

The magnetic-field profiles that emerge from the simulations are inevitably smooth on the grid scale. However, scattering of non-thermal particles is dominated by magnetic-field structures with length scale close to gyro-scale of the particles in question:
$\lambda \sim r_{\rm g} \sim E_\ur{PeV} B_{\mu {\rm G}} \,{\rm pc}$. Since the grid resolution in the vicinity of the WTS is of the order $0.1$\,pc, it is clear that a sub-grid model for self-excited magnetic field fluctuations is needed for particles with energies of interest in this work. In determining the maximum energy at each shock we have made the assumption that DSA operates at maximum efficiency. This entails the implicit assumption of Bohm diffusion, which demands the generation of self-excited MHD modes. To motivate this, we investigate which cosmic-ray driven instabilities can operate, and to what level we might expect the fields to grow/saturate. 

The two most commonly invoked processes for self-excitation of scattering fields for energetic cosmic rays are the resonant streaming instability \cite[e.g.][]{Wentzel} and the non-resonant hybrid instability \citep{Bell2004,Bell2005}. Note that the resonant instability (RSI) has largely only been studied in the context of quasi-parallel shocks, which as we demonstrated applies only in a limited range of the available shock surface. Meanwhile, the non-resonant instability (NRI) can operate at all possible obliquities, though only acts if the cosmic-ray currents can overcome the magnetic field tension at all resonant scales. This condition can be expressed as \citep{Bell2004,reville21}
\begin{equation}
 \chi=  \frac{P_\ur{cr}}{\rho u^2} \frac{u}{c} M_\ur{A}^2 \gg 1\, ,
\end{equation}
which is easily satisfied for the typical conditions found in our simulations; for $B_\ur{bg}=0.5\,\mu$G, $u=3000\kms$, and $n=0.05\pccm$, $\chi \sim 10^3 ({P_\ur{cr}}/{\rho u^2}) $.

Different estimates for the saturated magnetic field associated to each instability can be found in the literature  \cite[e.g.][]{BellLucek,amato2006, Bell2004,matthews}
\begin{equation}
    \frac{\delta B^2_{\rm sat}}{4\pi} \approx \left\lbrace \begin{array}{cc}
      2 \eta_\ur{cr}  M_\ur{A} \, \frac{B^2_\ur{bg}}{4\pi}    &  \enspace  \mbox{RSI} \\
     \eta_\ur{cr}\, \rho u^2 \,\frac{u}{c}     & \enspace \mbox{NRI}
    \end{array}
    \right. \, ,
\end{equation}
for a background field, $B_\ur{bg}$, and an acceleration efficiency $\eta_\ur{cr} = P_\ur{cr}/\rho u^2$, typically thought to be about 10\%.
Numerically we can express the above as 
\begin{equation}
    \delta B_\ur{sat}  \approx \left\lbrace \begin{array}{rc}
10 \left( \frac{B_\ur{bg}}{1\,\mu\ur{G}} \right)^{1/2} \left( \frac{u}{1000\kms} \right)^{1/2} \left( \frac{n}{1\pccm} \right)^{1/4} \left( \frac{\eta_\ur{cr}}{0.1} \right)^{1/2} \mu \ur{G} & \mbox{RSI} \\
8\left( \frac{u}{1000\,\kms} \right)^{3/2}  \left( \frac{n}{1\pccm} \right)^{1/2} \left( \frac{\eta_\ur{cr}}{0.1} \right)^{1/2}\,\mu\ur{G} & \mbox{NRI} 
    \end{array}
    \right. 
    \label{eq:res-str}
\end{equation}
Adopting typical values inferred from simulations, a saturated magnetic field approximately an order of magnitude larger than that of the background appears achievable. Although it is not expected that the Bohm scattering limit in the saturated field can be applied at all energy scales, the results do suggest that field amplification is possible, which lends confidence to our assumption that DSA is operating at, or close to, the maximal efficiency.

\subsection{Particle spectrum in the TeV range}
\label{sec:spec-model}

Considering the variation of parameters across the cluster WTS surface, in particular the variation in $B$, the spectrum of particles accelerated at the WTS is expected to have multiple components with different cut-offs. In Fig.~\ref{fig:spec}, we present a toy model for the spectrum expected from particle acceleration at sheet base and cone shocks. We add a powerlaw component for each pair of $M_\ur{S}$ and $B$ values in the immediate upstream extracted for each line-out. The powerlaws follow $\d N/\d E \propto E^{-\alpha}\exp(E/E_\ur{max})$, where the spectral index $\alpha = (q+1)/(q-1) $ is determined by the compression ratio, $q$, inferred from the upstream Mach number, 
\begin{equation}
	q = \frac{\gamma+1}{(\gamma-1)+2M_\ur{S}^{-2}} \, ,
\end{equation}
where $\gamma$ is the adiabatic index. The cut-off energy is set by $E_\ur{max}$, which in turn is set by $B$ along each individual line-out. The velocity along the shock normal and $d$ are chosen as in Eq.~\ref{eq:max-shb} and \ref{eq:max-con}. The velocity distribution was not incorporated into the model, because it is contaminated with downstream values and the change in direction of $\nabla M_\ur{S}$ in the numerically smoothed shock (see Sect.~\ref{sec:cl-acc}). $B\gtrsim3\,\mu$G is not sustained over regions ${\gtrsim}1\,$pc. We therefore set $d$ to 1\,pc if $B>3\,\mu$G.

The summed spectra for cones and sheet bases are shown in Fig.~\ref{fig:spec}. The spectra are steep, intersecting with $\d N/\d E \propto E^{-3}$ at ${\sim}1\,$PeV. The highest cut-off energy is ${\sim}500\,$TeV for both sheet bases and cones. Both spectra show significant curvature. All simulation runs show qualitatively similar results. Our toy model demonstrates that steep curved spectra, such as observed in $\gamma$-rays for multiple clusters \citep[e.g.][]{lhaaso24-w43}, can naturally arise from the inhomogeneous distribution of magnetic-field values at the cluster WTS. The curvature results from components with different cut-off energies. As discussed in Sect.~\ref{sec:cl-acc}, the cluster WTS has $M_\ur{S}\gtrsim5$ everywhere and therefore the spectral index is always close to 2.

\subsection{Particle transport}
\label{sec:transp}

In Sect.~\ref{sec:fldiff}, we inferred a diffusivity of $D = (\Delta x)^2/(2s)\sim0.5\mbox{--}2\,$pc for streamlines of the magnetic field, where $\Delta x$ is the linear spatial displacement and $s$ the length along the streamline. Non-thermal particles, to lowest order, follow the large-scale field, performing helical motion. In the transonic sheets, the field shows a quasi-radial morphology (see Fig.~\ref{fig:strln}). This may lead to fast escape of particles if the scattering rate is low. Since constraining turbulence on particle gyro-scales is beyond the capabilities of our approach, this remains an unknown. The cones show a more diverse magnetic field morphology, which includes regions where the fieldlines appear highly tangled, sections of quasi-perpendicular fieldlines in the cluster WTS downstream where the field is amplified, and loops extending outward on various scales. Around the shock, $\v{B}$ shows a wide range of obliquities (see Fig.~\ref{fig:upstr-vals}, right-most panel). While perpendicular fields in the shock downstream can in principle confine particles close to the shock, it seems likely that any such effect is minor, since particles could still escape through the transonic sheets.

As expected in ideal MHD, fieldlines are anchored in the core from which the originate. This leads to quasi-perpendicular fields at the contact discontinuity at the edge of the superbubble, which could effectively confine particles inside the superbubble. In realistic scenarios, the shell is likely to at least partly break apart due to instabilities \citep[e.g.][]{elbadry2019,Lancaster2024}. However, this phenomenon is not well constrained, among other factors because the magnetic field can stabilise the shell \citep[see][]{chandrasekhar61}. Our simulations are tailored to studying the cluster core and wind and do not have sufficient resolution at the shell to draw meaningful conclusions.

\section{Conclusion}
\label{sec:concl}

We have performed 3D ideal-MHD simulations of stellar winds around compact, young, massive star clusters. Prominent examples of such objects include Westerlund~1, 30~Doradus~C, and R136. We model 46 individual stars with $M>40\,\Msun$ and inject winds kinetically to study the 3D flow and magnetic field geometry on the superbubble scale. We include the effects of WR winds and study clusters with two different radii ($R_\ur{c}=0.6\,$pc and 1\,pc) and discuss three different scenarios for the magnetic field. Our main conclusions are:

\begin{enumerate}
    \item The flow shows an intricate morphology, which significantly deviates from spherical symmetry. We recover the subsonic cluster core, supersonic collective cluster wind, and tenuous superbubble interior medium expected from 1D theory. The position and power of individual stars critically influences the morphology of the flow. The cluster wind has non-uniform properties and is locally impacted by downstream material from individual stellar winds. For example, regions in the flow path of magnetic stars have an above-average magnetic field. 
    \item In addition to the three zones mentioned above, we find firstly that strong stellar winds can extend from the core all the way to the cluster WTS (``coupled winds''). Secondly, we find transonic sheets extending well beyond the cluster WTS with an internal structure that resembles that of the ``shock diamond'' phenomenon. 
    \item The cluster WTS has a Mach number of $M_\ur{S}\sim 10$ at the end of the simulation (390\,kyr) for a cluster radius of 0.6\,pc and $M_\ur{S}\sim 6$ for 1\,pc. Since the typical age of young compact clusters exceeds the simulation time, we conclude that compact clusters ($R_\ur{c}<2\mbox{--}3\,$pc) can in general be expected to have a strong cluster WTS. The cluster WTS is found to be non-spherical and smaller in volume compared to 1D theory. The cluster WTS can be sub-divided into regions of quasi-parallel flow and highly oblique flow (``sheet bases'' and ``cones'', respectively).
    \item The magnetic field morphology is characterised by loops of various sizes, which stay anchored in the cluster core, as well as coherent, quasi-radial field-line bundles extending outward preferentially inside transonic sheets. In the core, the magnetic field is highly tangled and locally dominated by Parker spiral fields from individual stars. The obliquity of the magnetic field varies over the cluster WTS surface. Mixing of flow from magnetic stars into the bulk medium contributes significantly to the observed magnetisation of the bulk flow. Across all runs, we find median magnetic field magnitude, $B$, of $8\mbox{--}35\,\mu$G in the core and $4\mbox{--}20\,\mu$G in the superbubble. Values of up to ${\sim}1\,$mG are reached in the vicinity of magnetic stars. These findings are consistent with previous work simulating only the cluster core \citep{badmaev22}. The magnetic field is found to scale slightly steeper that $r^{-1}$ in the cluster wind. 
    \item We discuss particle acceleration and propagation around young, compact star clusters. We find a large spread in $B$ in the cluster WTS upstream, which ultimately leads to the prediction of steep particle spectra in the TeV energy range (${\propto}E^{-3}$ with curvature). This finding is in line with recent $\gamma$-ray observations of young clusters \citep[e.g.][]{lhaaso24-w43}. However, we deem the scenario of PeV particle acceleration at the cluster WTS unlikely, as the Hillas limit barely reaches 200\,TeV in the most highly magnetised regions. Dynamo effects on scales ${<}0.01\,$pc would have to increase the magnitude of $B$ by a factor of at least five over that observed in the simulation to reach PeV energies. Streaming instabilities can in principle amplify the magnetic field above the large-scale value from the simulation, but whether they reach saturation on the relevant scales is uncertain.
\end{enumerate}
Our work show-cases the non-trivial nature of stellar-wind interaction around young star clusters. Several parameters, such as the distribution and median of $B$ at the cluster WTS, depend on the spatial distribution of stars. Considering that the population of young star clusters detected in $\gamma$-rays is highly heterogeneous in terms of age, compactness, ambient medium, and stellar population, it appears challenging to develop a unified model of particle acceleration for these regions. Physical models of $\gamma$-ray emission require an understanding of the stellar population and its history and feedback onto the environment. Nevertheless, the average properties discussed in this work can serve a purpose for order of magnitude estimates and give a qualitative impression of the nature of collective effects. A multi-wavelength effort is required to gain further insight into stellar-wind interaction and magnetic fields in star-cluster environments. Future $\gamma$-ray observations with LHAASO and CTAO will further characterise the cosmic ray populations in the vicinity of young star clusters and provide valuable constraints for acceleration models. X-ray and radio data can constrain the magnetic field to investigate the plausibility of acceleration to high energies. Infrared studies, in particular at high resolution as facilitated by JWST, can illuminate ambient gas and dust structures, constraining the nature of feedback processes.

\begin{acknowledgements}
      The simulations were carried out at the Max-Planck Computing and Data Facility (MPCDF). We thank the developers of \texttt{PLUTO} and the analysis tool \texttt{pyPLUTO} \citep{mattia25}, which have facilitated this work. We acknowledge insightful discussion with A. C. Sander, J. S. Wang, and G. Mattia and are grateful for the constructive feedback from the referee. We thank the organisers and attendees of the TOSCA meeting in autumn 2024\footnote{\url{https://indico.ict.inaf.it/event/2878/overview}} for interdisciplinary discussion. 
\end{acknowledgements}

%
%

\bibliography{MHD1}
\bibliographystyle{aa}



\begin{appendix}
\section{Comparison to 1D theory}
\label{app:1d}

\begin{figure}
	\centering
	\includegraphics[width=\linewidth]{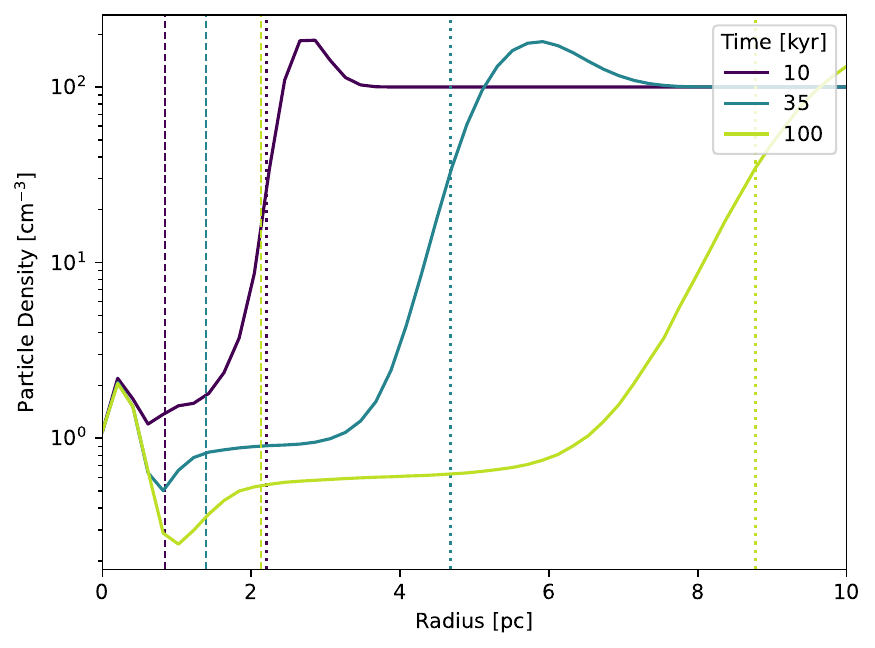}
	\caption{Radial density profile of the simulated superbubble for cluster I in the initial 100\,kyr of the simulation. The dashed and dotted lines indicate the cluster-wind termination shock and bubble radii according to the spherically symmetric, analytic model by \citet{weaver77}.}
	\label{fig:rho}
\end{figure}

\begin{figure*}
    \centering
    \includegraphics[width=0.8\linewidth]{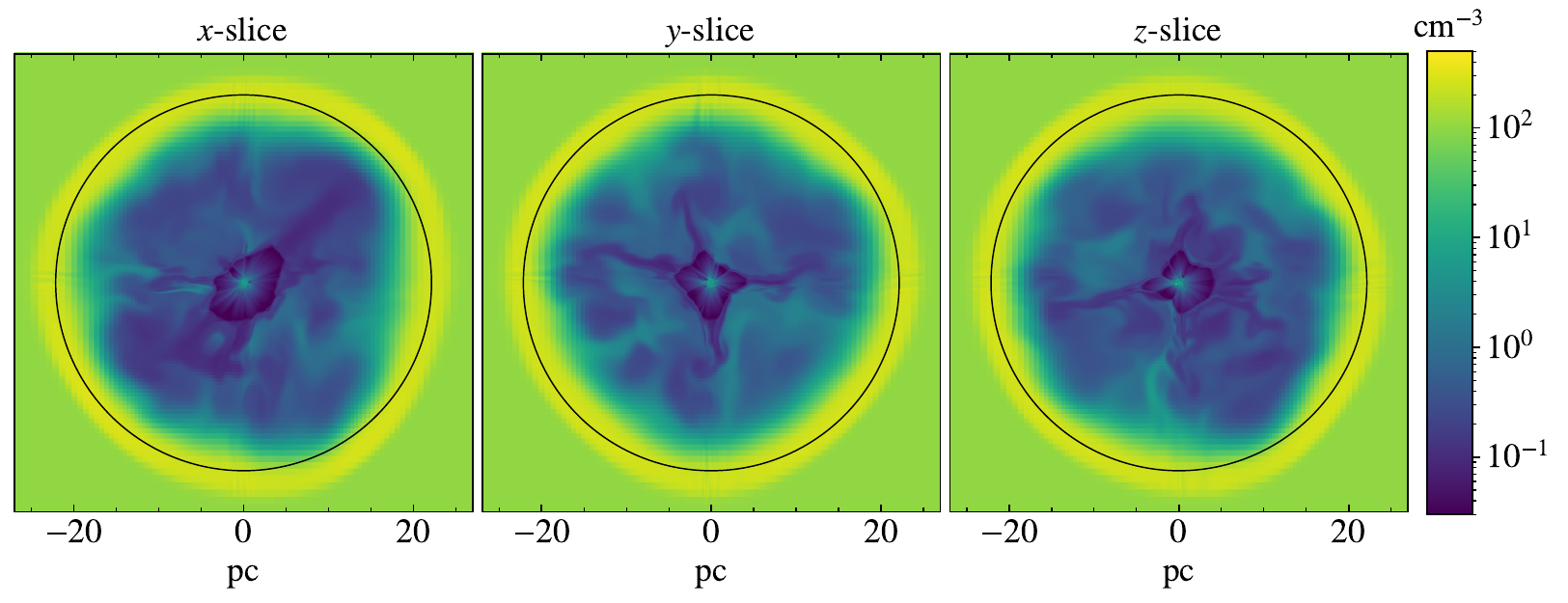}
    \includegraphics[width=0.8\linewidth]{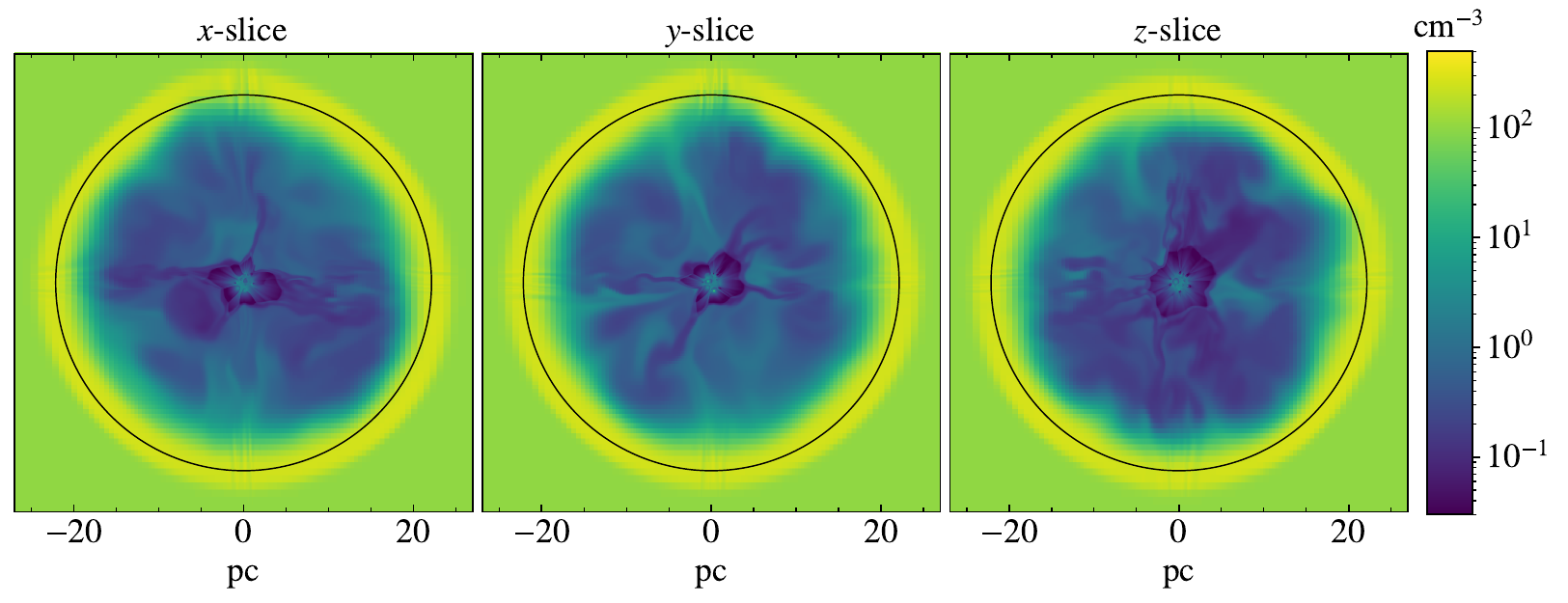}
    \caption{Density slices of cluster I (top) and III (bottom) showing the full simulation domain at 390\,kyr. Black circles indicate the superbubble radius predicted by \citet{weaver77}.}
    \label{fig:full-box}
\end{figure*}

\begin{figure}
	\centering
	\includegraphics[width=\linewidth]{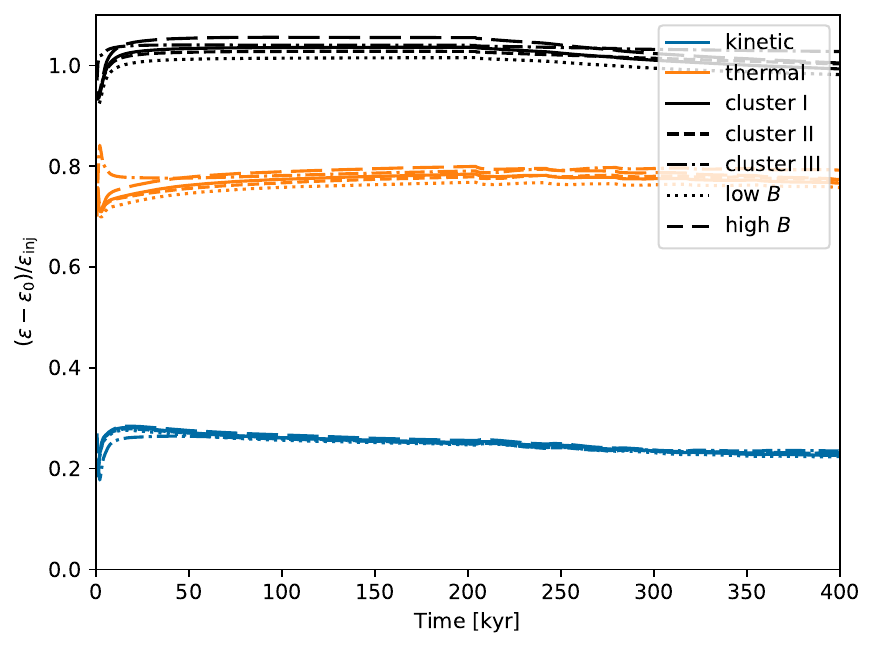}
	\caption{Kinetic and thermal energy in the full simulation domain relative to the total injected energy, $\epsilon_\ur{inj}=L_\ur{w}t$, for different simulations. The initial energy, $\epsilon_0$, is subtracted. Magnetic energy is not shown but is below the percent level.}
	\label{fig:energies}
\end{figure}


This section compares the simulation results to 1D theory \citep{weaver77}. Figure \ref{fig:rho} shows the density as a function of radius in the first 100\,kyr of the simulation. The familiar bubble structure for a compact cluster is recovered: a decrease in the density in the free wind is followed by an increase behind the cluster WTS, which in turn is followed by a zone of constant density in the superbubble interior. Then, the density rises sharply and beyond the ambient value at the shell. While the prediction by \citet{weaver77} for the bubble radius agrees with the position of the shell, the termination shock radius is overestimated, due to multi-dimensional effects (see Sect.~\ref{sec:evo}). Figure~\ref{fig:full-box} shows density slices of the full simulation domain at 390\,kyr, highlighting that the position of the forward shock is consistent with 1D theory over the full simulation run time. Figure~\ref{fig:energies} shows the thermal, kinetic, and total energy in the simulation domain for all runs, relative to the injected wind power. This agrees with the \citet{weaver77} which predicts that, when cooling is neglected, ${\sim}22\%$ of the injected energy goes into kinetic energy and ${\sim}78\%$ into thermal energy.

\section{Star clusters}
\label{app:tab-stars}

\begin{table*}
    \centering
    \begin{tabular}{c|c|c|c|c|c|c|c}
        Nr.& $M$ & $T_\ur{eff}$ &  $R$ & $t_\ur{MS}$ & $\dot{M}$ & $u_\infty$ & $\Lw$  \\
        \hline
         & $\ur{M}_\odot$ & K &  $\ur{R}_\odot$ & kyr & $10^{-6}\Msyr$ & km/s & $10^{36}\ergs$ \\
        \hline
        \hline
        $1^{\ur{a}, \ur{b}}$ & 40.2 & 45{,}519 & 11.4 & 4591 & 0.74 & 2632 & 1.61 \\
        2 & 40.7 & 45{,}708 & 11.5 & 4548 & 0.77 & 2635 & 1.69 \\
        3 & 41.2 & 45{,}899 & 11.6 & 4506 & 0.81 & 2637 & 1.77 \\
        4 & 41.7 & 46{,}094 & 11.7 & 4463 & 0.85 & 2640 & 1.86 \\
        $5^\ur{b}$ & 42.3 & 46{,}291 & 11.8 & 4421 & 0.89 & 2643 & 1.95 \\
        6 & 42.8 & 46{,}491 & 12.0 & 4379 & 0.93 & 2647 & 2.05 \\
        7 & 43.4 & 46{,}694 & 12.1 & 4337 & 0.97 & 2650 & 2.15 \\
        8 & 44.0 & 46{,}901 & 12.2 & 4295 & 1.02 & 2653 & 2.26 \\
        9 & 44.6 & 47{,}111 & 12.3 & 4254 & 1.07 & 2656 & 2.38 \\
        $10^\ur{b}$ & 45.2 & 47{,}324 & 12.5 & 4212 & 1.12 & 2659 & 2.50 \\
        $11^{\ur{a}}$ & 45.9 & 47{,}541 & 12.6 & 4170 & 1.17 & 2663 & 2.62 \\
        12 & 46.5 & 47{,}761 & 12.8 & 4129 & 1.23 & 2666 & 2.76 \\
        13 & 47.2 & 47{,}984 & 12.9 & 4088 & 1.29 & 2670 & 2.90 \\
        $14^\ur{b}$ & 47.9 & 48{,}212 & 13.1 & 4047 & 1.36 & 2673 & 3.06 \\
        15 & 48.6 & 48{,}443 & 13.2 & 4006 & 1.43 & 2677 & 3.22 \\
        16 & 49.4 & 48{,}679 & 13.4 & 3965 & 1.50 & 2681 & 3.39 \\
        17 & 50.2 & 48{,}919 & 13.6 & 3924 & 1.57 & 2685 & 3.57 \\
        18 & 51.0 & 49{,}163 & 13.7 & 3883 & 1.65 & 2688 & 3.76 \\
        $19^\ur{b}$ & 51.8 & 49{,}411 & 13.9 & 3843 & 1.74 & 2692 & 3.96 \\
        20 & 52.7 & 49{,}664 & 14.1 & 3803 & 1.83 & 2696 & 4.18 \\
        $21^{\ur{a}}$ & 53.6 & 49{,}921 & 14.3 & 3762 & 1.92 & 2701 & 4.41 \\
        22 & 54.6 & 50{,}184 & 14.5 & 3722 & 2.02 & 2705 & 4.65 \\
        23 & 55.6 & 50{,}451 & 14.8 & 3682 & 2.12 & 2709 & 4.91 \\
        $24^\ur{b}$ & 56.6 & 50{,}724 & 15.0 & 3642 & 2.23 & 2714 & 5.18 \\
        25 & 57.7 & 51{,}003 & 15.2 & 3602 & 2.35 & 2718 & 5.47 \\
        26 & 58.8 & 51{,}286 & 15.5 & 3563 & 2.48 & 2723 & 5.78 \\
        27 & 60.0 & 51{,}576 & 15.7 & 3523 & 2.61 & 2728 & 6.11 \\
        $28^\ur{b}$ & 61.2 & 51{,}872 & 16.0 & 3484 & 2.74 & 2733 & 6.46 \\
        29 & 62.5 & 52{,}174 & 16.3 & 3445 & 2.89 & 2738 & 6.82 \\
        30 & 63.8 & 52{,}482 & 16.6 & 3406 & 3.04 & 2743 & 7.21 \\
        $31^{\ur{a}}$ & 65.3 & 52{,}798 & 17.0 & 3367 & 3.20 & 2749 & 7.63 \\
        $32^\ur{b}$ & 66.8 & 53{,}120 & 17.3 & 3328 & 3.37 & 2754 & 8.06 \\
        33 & 68.3 & 53{,}450 & 17.7 & 3289 & 3.55 & 2760 & 8.53 \\
        34 & 70.0 & 53{,}787 & 18.1 & 3250 & 3.74 & 2766 & 9.02 \\
        35 & 71.7 & 54{,}133 & 18.5 & 3212 & 3.94 & 2772 & 9.54 \\
        36 & 73.6 & 54{,}487 & 18.9 & 3174 & 4.15 & 2779 & 10.08 \\
        37 & 75.6 & 54{,}849 & 19.4 & 3135 & 4.36 & 2785 & 10.66 \\
        38 & 77.7 & 55{,}220 & 19.9 & 3097 & 4.59 & 2792 & 11.27 \\
        $39^ \ur{b}$ & 79.9 & 55{,}601 & 20.4 & 3059 & 4.82 & 2799 & 11.91 \\
        40 & 82.3 & 55{,}992 & 21.0 & 3022 & 5.07 & 2807 & 12.58 \\
        $41^{\ur{a}, \ur{b}}$ & 84.8 & 56{,}393 & 21.6 & 2984 & 5.32 & 2814 & 13.28 \\
        42 & 87.5 & 56{,}805 & 22.3 & 2947 & 5.58 & 2822 & 14.00 \\
        43 & 90.4 & 57{,}227 & 23.0 & 2909 & 5.85 & 2831 & 14.76 \\
        44 & 93.6 & 57{,}662 & 23.8 & 2872 & 6.12 & 2839 & 15.54 \\
        45 & 97.0 & 58{,}109 & 24.7 & 2835 & 6.39 & 2849 & 16.33 \\
        46 & 100.7 & 58{,}568 & 25.7 & 2798 & 6.66 & 2858 & 17.14 \\
    
    \end{tabular}
    \caption{Parameters of the stars in our model cluster, showing stellar mass, $M$, effective temperature, $T_\ur{eff}$, radius, $R$, main sequence life-time, $t_\ur{MS}$, mass-loss rate, $\dot{M}$, terminal wind velocity, $u_\infty$, and wind power, $\Lw=0.5\dot{M}u_\infty^2$. Stars with superscript a are magnetic stars. Their dipole field strength is $B_\ur{s}=1\,\ur{k}$G in runs with standard magnetic field (clusters I--III in Tab.~\ref{tab:sims}) and $100\,$G in low $B$ case. In the high $B$ case, the number of magnetic stars is doubled, but individual stars still have $B_\ur{s}=1\,\ur{k}$G (superscript b). In the simulation, the main sequence phase is shortened. The most massive star (Nr.~46) enters the Wolf-Rayet phase at 200\,kyr, followed by stars Nr.~45--42, enter with a time delay according to their difference in life-time to star Nr.~46.}
    \label{tab:stars}
\end{table*}

\begin{figure*}
	\centering
	\begin{subfigure}{0.33\linewidth}
		\includegraphics[width=\linewidth]{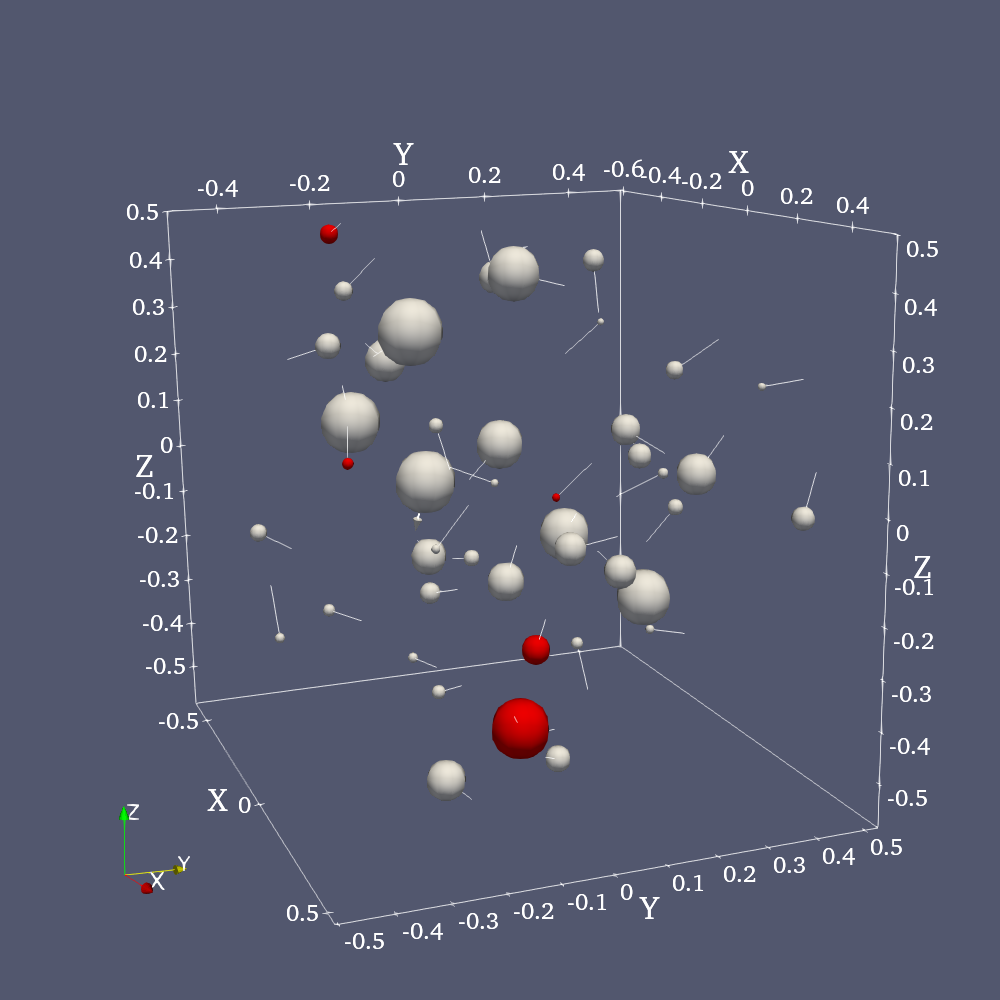}
		\caption{cluster I, $R_\ur{c}=0.6\,$pc}
	\end{subfigure}
	\begin{subfigure}{0.33\linewidth}
		\includegraphics[width=\linewidth]{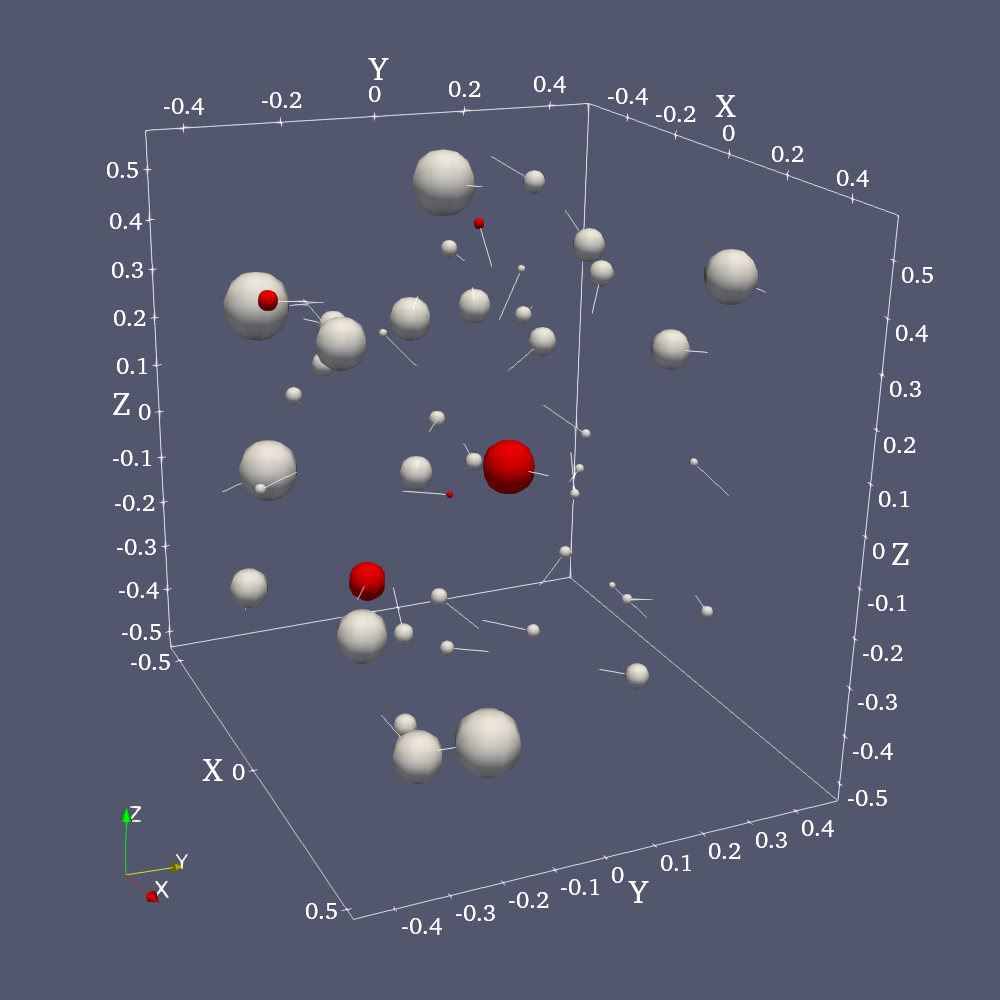}
		\caption{cluster II, $R_\ur{c}=0.6\,$pc}
	\end{subfigure}
	\begin{subfigure}{0.33\linewidth}
		\includegraphics[width=\linewidth]{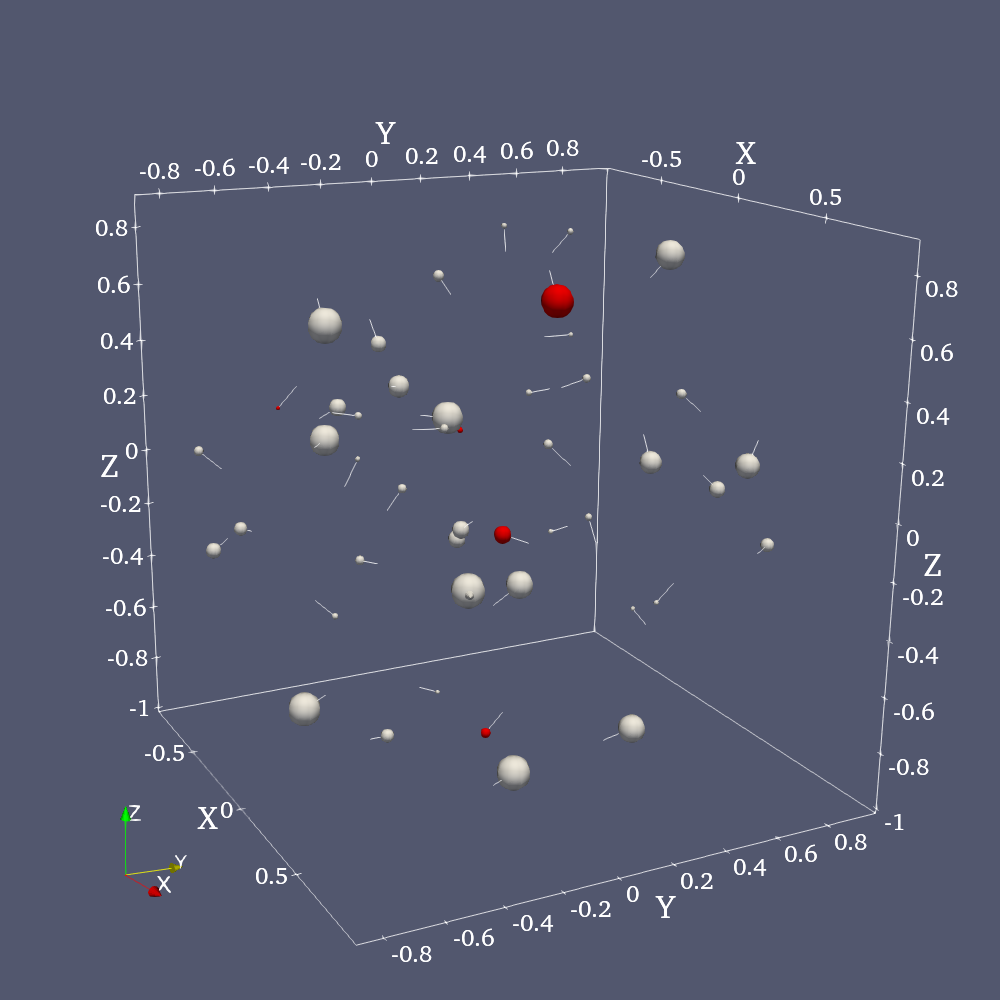}
		\caption{cluster III, $R_\ur{c}=1\,$pc}
	\end{subfigure}
	
	\caption{3D renderings of the three different model clusters analysed in this work. All clusters have the same stellar content listed in Tab.~\ref{tab:stars}. The size of the sphere is scaled by wind power. Magnetic stars are marked in red. The lines indicate the orientation of the magnetic axes. Clusters I and II have the same compactness $R_\ur{c}=0.6\,$pc and only differ in the spatial distribution of stars. Most notably, a rather powerful magnetic star is located close to the centre in cluster II, while in cluster I, it is at the outskirts. Cluster III is less compact, $R_\ur{c}=1\,$pc.}
	\label{fig:cluster3d}
\end{figure*}

In Tab.~\ref{tab:stars}, we provide the stellar content of our model cluster. For a discussion on how the parameters where selected, see Sect.~\ref{sec:setup}. Figure~\ref{fig:cluster3d} shows 3D renderings of cluster I, II, and III.

\section{Supplementary plots on the magnetic field}
\label{app:b-suppl}

\begin{figure*}
	\centering
	\includegraphics[width=0.242\linewidth]{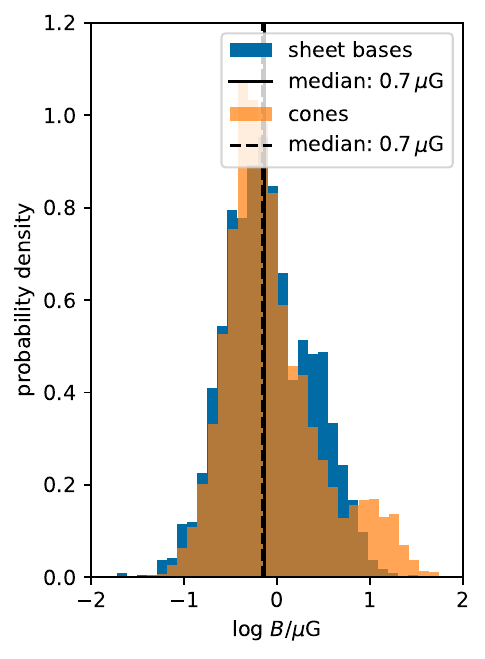}
	\includegraphics[width=0.242\linewidth]{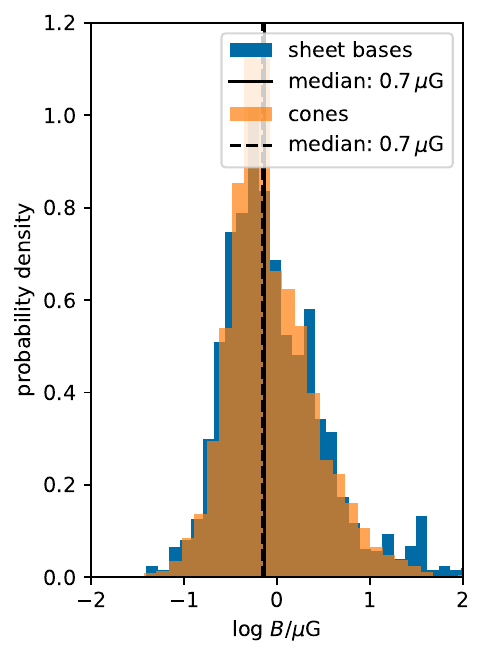}
	\includegraphics[width=0.242\linewidth]{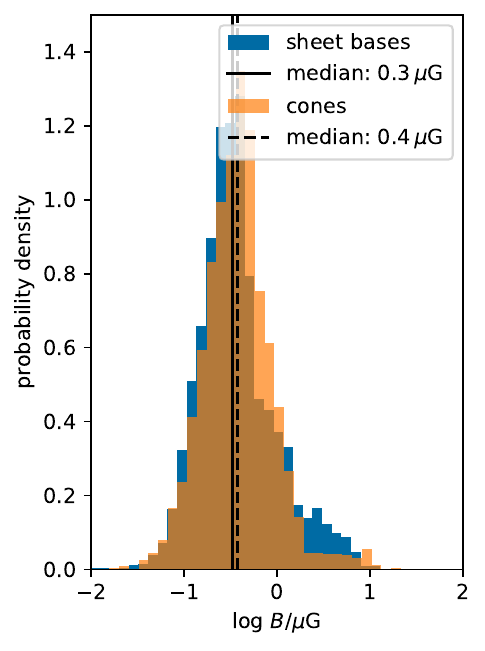}
	\includegraphics[width=0.242\linewidth]{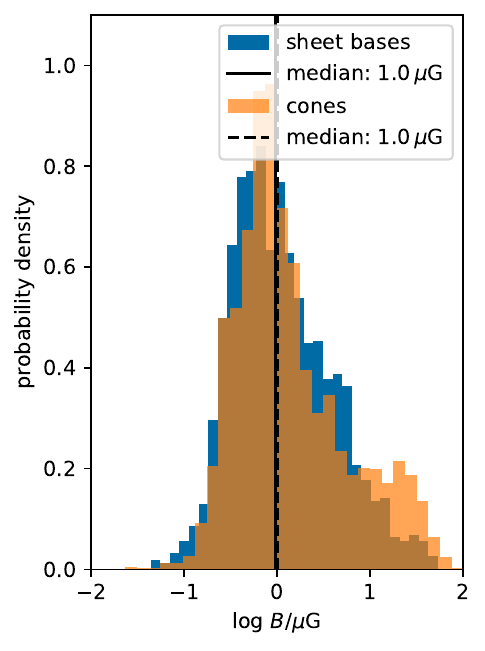}
	\caption{Distribution of magnetic field magnitude in the immediate cluster-wind termination shock upstream clusters III, II, the low $B$ cluster, and the high $B$ cluster (from left to right). Note that cluster II shows fewer high $B$ outliers than cluster I (see Fig.~\ref{fig:upstr-vals} in the main text). Both clusters differ only differ in the spatial distribution of stars (see Fig.~\ref{fig:cluster3d}). For further context, see Sect.~\ref{sec:cl-acc}.}
	\label{fig:bupstr-other-sims}
\end{figure*}

\begin{figure*}
    \centering
    \includegraphics[width=0.45\linewidth]{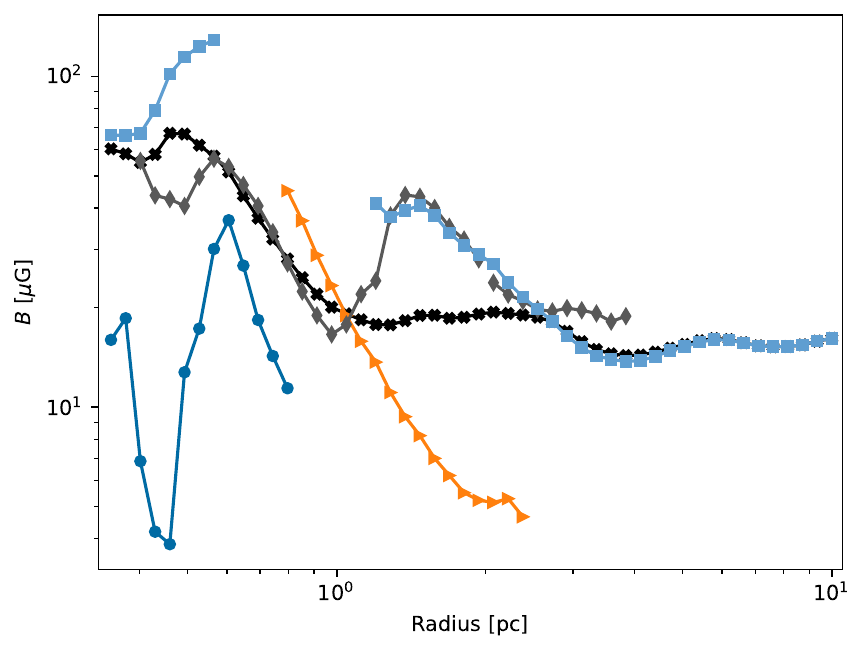}
    \includegraphics[width=0.45\linewidth]{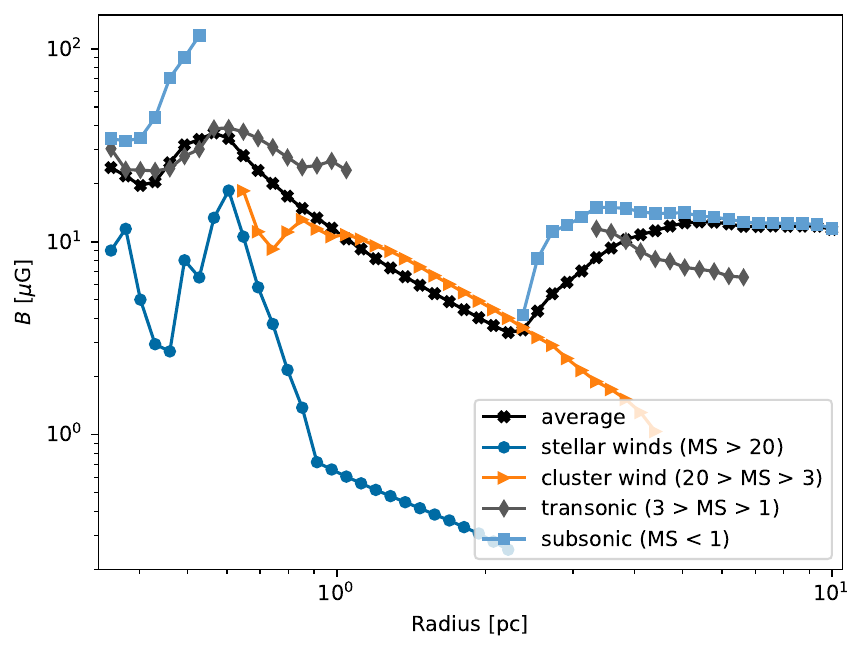}
    \caption{Radial profiles of the mean magnitude of $\v{B}$ for cluster I in different regions. The left panel shows radial profiles at 200\,kyr and the right at 390\,kyr. For further context, see Sect.~\ref{sec:rprofile}.}
    \label{fig:b-rad-44}
\end{figure*}

\begin{figure*}
    \centering
    \includegraphics[width=0.45\linewidth]{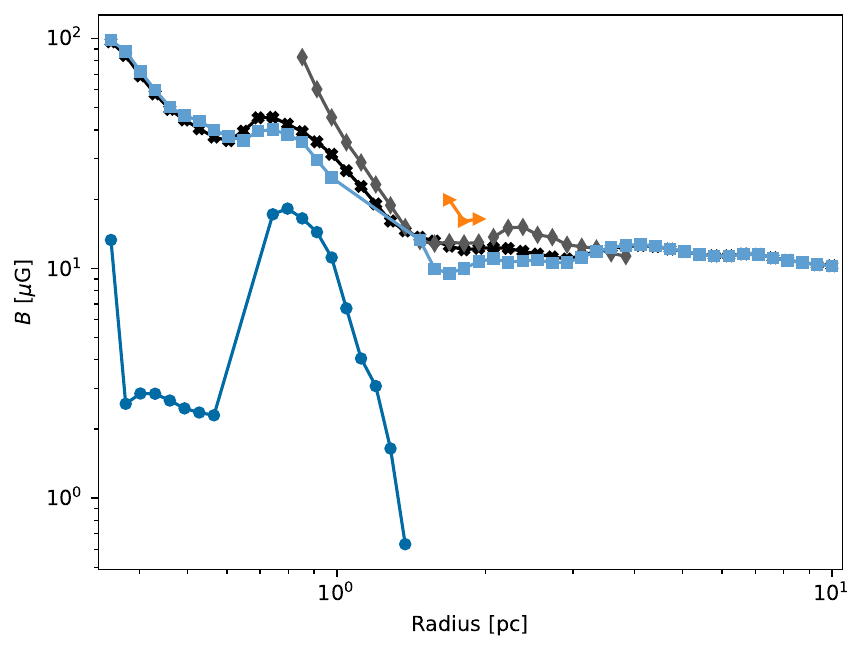}
    \includegraphics[width=0.45\linewidth]{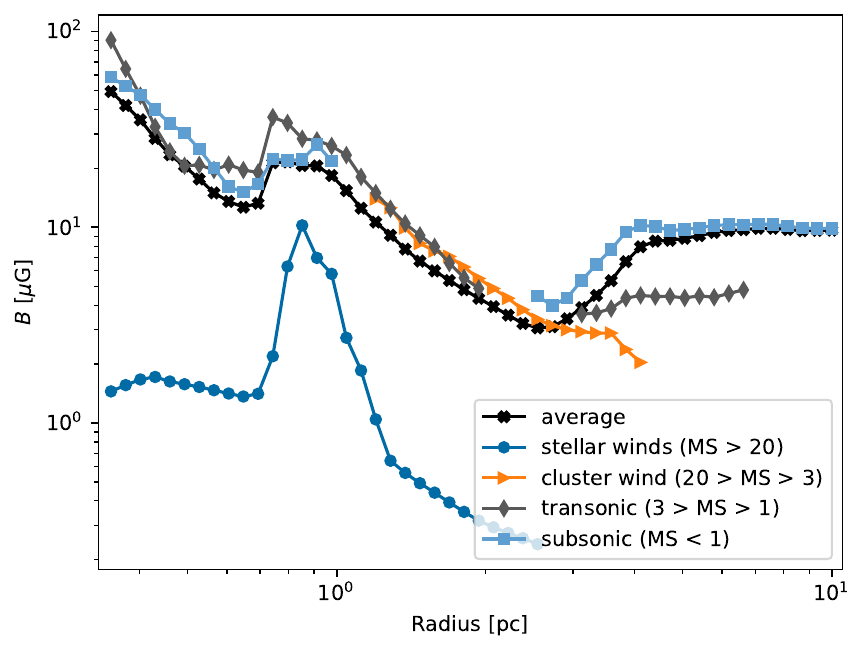}
    \caption{Same as Fig.~\ref{fig:b-rad-44} for cluster III.}
    \label{fig:b-rad-45}
\end{figure*}

\begin{figure*}
    \centering
    \includegraphics[width=0.45\linewidth]{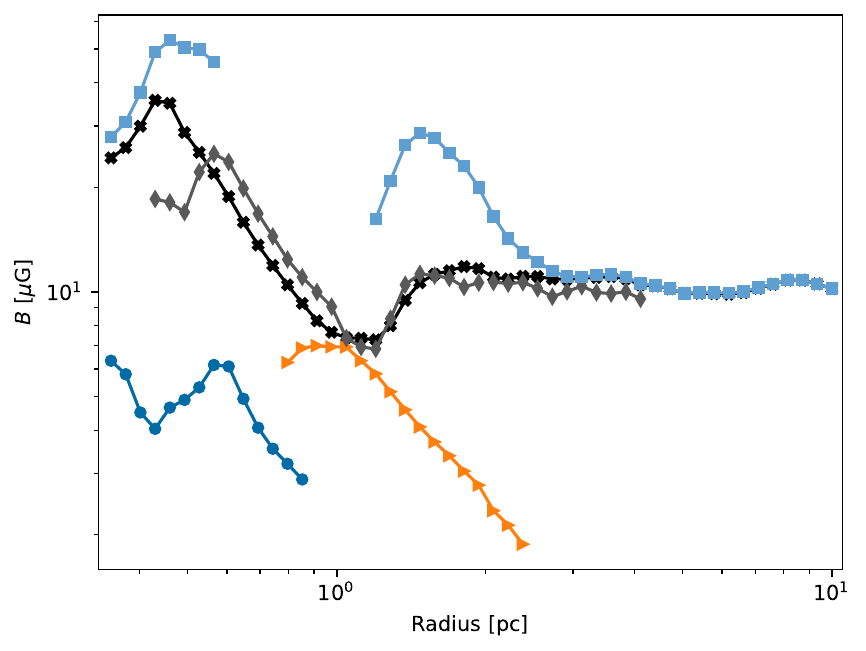}
    \includegraphics[width=0.45\linewidth]{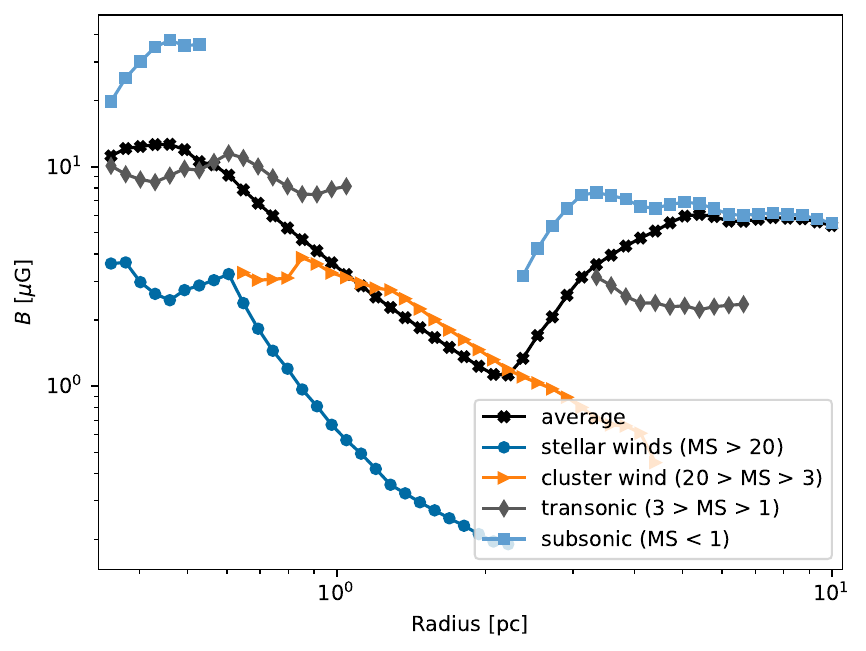}
    \caption{Same as Fig.~\ref{fig:b-rad-44} for the low $B$ cluster.}
    \label{fig:b-rad-46}
\end{figure*}

\begin{figure*}
    \centering
    \includegraphics[width=0.45\linewidth]{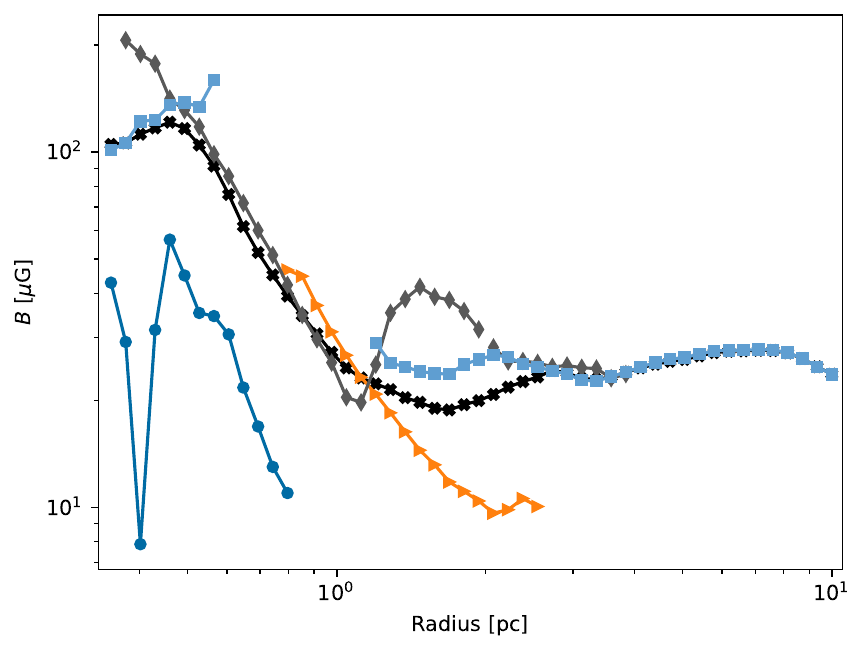}
    \includegraphics[width=0.45\linewidth]{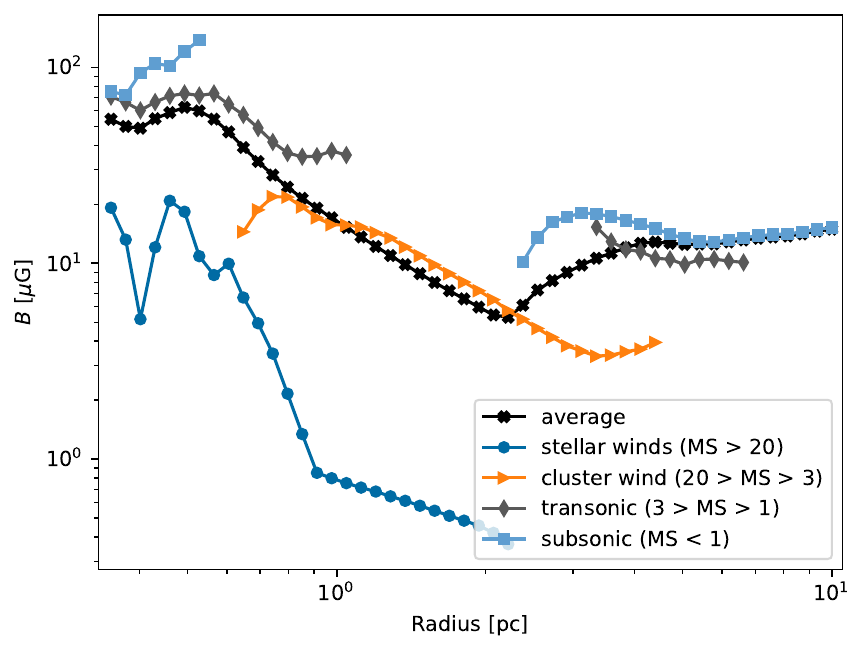}
    \caption{Same as Fig.~\ref{fig:b-rad-44} for the high $B$ cluster.}
    \label{fig:b-rad-47}
\end{figure*}


Figure~\ref{fig:bupstr-other-sims} shows the upstream distributions of $B$ for different simulation runs. Note that cluster II shows significantly fewer high $B$ outliers than cluster I, which is discussed in the main text. This is due to the more central location of the magnetic star in the star cluster for cluster II.

Figures~\ref{fig:b-rad-44}--\ref{fig:b-rad-47} show radial profiles of the mean magnitude of $\v{B}$ for the simulation runs not included in the main text. The same general trends can be observed across all profiles. Note, however, the small extension of the cluster wind ($20>M_\ur{S}>3$) for the cluster with $R_\ur{c}=1\,$pc instead of 0.6\,pc (cluster III, Fig.~\ref{fig:b-rad-45}) at 200\,kyr. This demonstrates that the cluster wind requires a longer time to reach a given Mach number for less compact clusters. Even at 390\,kyr, the transonic component of the cluster wind ($3>M_\ur{S}>1$) is ${>}10\%$ for most radii.

\end{appendix}
\end{document}